\documentclass[aps,showpacs]{revtex4}
\usepackage{amsmath}
\usepackage{graphicx}
\begin{document}


\title{{\it Ab initio} shell model with a genuine three-nucleon force 
for the $p$-shell nuclei}
\author{Petr Navr\'atil}
\affiliation{Lawrence Livermore National Laboratory, L-414, P.O. Box 808, 
Livermore, CA  94551}
\author{W. Erich Ormand}
\affiliation{Lawrence Livermore National Laboratory, L-414, P.O. Box 808, 
Livermore, CA  94551}

\begin{abstract}
The {\it ab initio} no-core shell model (NCSM) is extended to include
a realistic three-body interaction in calculations for $p$-shell nuclei.
The NCSM formalism is reviewed and new features needed in calculations
with three-body forces are discussed in detail.
We present results of first applications to $^{6,7}$Li, $^6$He, $^{7,8,10}$Be, $^{10,11,12}$B,
$^{12}$N and $^{10,11,12,13}$C using the Argonne V8$^\prime$ nucleon-nucleon (NN) potential and 
the Tucson-Melbourne TM$^\prime$(99) three-nucleon interaction (TNI). In 
addition to increasing the total binding energy, we observe a substantial sensitivity 
in the low-lying spectra to the presence of the realistic three-body force and 
an overall improvement in the level-ordering and level-spacing in comparison to 
experiment. The greatest sensitivity occurs for states where the spin-orbit 
interaction strength is known to play a role. In particular, with the TNI
we obtain the correct ground-state spin for $^{10,11}$B and $^{12}$N, $^{12}$B, 
contrary to calculations with NN potentials only. 
\end{abstract}
\maketitle

\section{Introduction}
The {\it ab initio} no-core shell model (NCSM) \cite{C12_NCSM} is a method to solve
the nuclear structure problem for light nuclei considered as systems
of $A$ nucleons interacting by realistic inter-nucleon forces. The calculations
are performed using a large but finite harmonic-oscillator (HO) basis. Due
to the basis truncation, it is necessary to derive an effective interaction
from the underlying inter-nucleon interaction that is appropriate for the basis 
employed. The effective interaction contains, in general, up to $A$-body 
components even if the underlying interaction had, e.g. only two-body terms.
In practice, the effective interaction is derived in a sub-cluster approximation
retaining just two- or three-body terms. A crucial feature of the method is its 
convergence to exact solution  with increasing basis size and/or an increase in
the effective interaction clustering.

In the past, applications were limited to only realistic two-nucleon
interactions. However, the recent introduction of the capability to derive a three-body
effective interaction \cite{fourb_NCSM} and apply it in either relative-coordinate 
\cite{Jacobi_NCSM} or Cartesian-coordinate \cite{v3eff}
formalism together with the ability to solve a three-nucleon system with a genuine 
three-nucleon force in the NCSM approach \cite{NCSM_TM} opens the possibility to include 
a realistic three-nucleon interaction (TNI) in the NCSM Hamiltonian and perform 
calculations for the $p$-shell nuclei.

We note that there are several methods that can be used to solve the $A=3,4$ systems
with realistic Hamiltonians that include a realistic three-body interaction.
However, until now, only  the Green's function Monte Carlo (GFMC) method 
\cite{av8p,wiringa00,pieper01,GFMC_9_10}
is capable to obtain solution for light $p$-shell nuclei with a Hamiltonian
that includes both realistic two- and three-nucleon force.

In this paper, we introduce an extension of the NCSM formalism to accommodate
the TNI and present first applications for several $p$-shell nuclei. The main purpose
of this paper is to introduce the formalism needed to perform {\it ab initio} 
shell-model calculations for the $p$-shell nuclei with Hamiltonians that include 
a realistic three-nucleon interaction. At the same time, we present a snapshot of
first applications
for several $p$-shell nuclei of different masses with the primary goal of
assessing a general impact of a realistic TNI on the structure
of different $p$-shell nuclei. In this study, we limit ourselves
to the use of a single TNI, the chiral-symmetry based Tucson-Melbourne 
TM$^\prime$(99) \cite{TMprime99}, combined with the Argonne V8$^\prime$ NN 
potential \cite{av8p}.
More detailed studies using a broader variety of realistic three-nucleon interactions
will follow in the future.    

In Section \ref{sec_theory}, the extension of the NCSM formalism to include
realistic three-nucleon forces is discussed. In Section \ref{sec_res},
we present first results obtained with Hamiltonians that include the Tucson-Melbourne
TM$^\prime$(99) TNI for $^{6,7}$Li, $^6$He, $^{7,8,10}$Be, $^{10,11,12}$B, $^{12}$N 
and $^{10,11,12,13}$C.
Some overall observations of the effect of the TNI are gathered in Section~\ref{TNI}.
Finally, we summarize our conclusions in Section \ref{sec_concl}.

\section{{\it Ab initio} no-core shell model with a three-nucleon force}\label{sec_theory}

A detailed description of the NCSM approach was presented, e.g. 
in Refs. \cite{C12_NCSM,fourb_NCSM,Jacobi_NCSM}.
Here, we emphasize extensions and modifications needed when a genuine
TNI is included. In the case when the TNI is considered, the starting Hamiltonian is
\begin{equation}\label{ham}
H_A= 
\frac{1}{A}\sum_{i<j}\frac{(\vec{p}_i-\vec{p}_j)^2}{2m}
+ \sum_{i<j}^A V_{{\rm NN}, ij} + \sum_{i<j<k}^A V_{{\rm NNN}, ijk} \; ,
\end{equation}
where $m$ is the nucleon mass, $V_{{\rm NN}, ij}$,
the NN interaction with both strong and electromagnetic components, 
$V_{{\rm NNN}, ijk}$ the three-nucleon interaction. In the NCSM, we employ a large
but finite harmonic-oscillator (HO) basis. Due to properties of the realistic nuclear 
interaction in Eq. (\ref{ham}),
we must derive an effective interaction appropriate for the basis truncation.
To facilitate the derivation of the effective interaction, we modify the
Hamiltonian (\ref{ham}) by adding to it the center-of-mass (CM) HO Hamiltonian
$H_{\rm CM}=T_{\rm CM}+ U_{\rm CM}$, where
$U_{\rm CM}=\frac{1}{2}Am\Omega^2 \vec{R}^2$,
$\vec{R}=\frac{1}{A}\sum_{i=1}^{A}\vec{r}_i$.
The effect of the HO CM Hamiltonian will later be subtracted
out in the final many-body calculation. Due to the translational invariance of the
Hamiltonian (\ref{ham}) the HO CM Hamiltonian has in fact no effect on the intrinsic
properties of the system in the infinite basis space. 
The modified Hamiltonian can be cast into the form
\begin{eqnarray}\label{hamomega}
H_A^\Omega &=& H_A + H_{\rm CM}=\sum_{i=1}^A h_i + \sum_{i<j}^A V_{ij}^{\Omega,A}
+\sum_{i<j<k}^A V_{{\rm NNN}, ijk} \nonumber \\ 
&=& \sum_{i=1}^A \left[ \frac{\vec{p}_i^2}{2m}
+\frac{1}{2}m\Omega^2 \vec{r}^2_i
\right] + \sum_{i<j}^A \left[ V_{{\rm NN}, ij}
-\frac{m\Omega^2}{2A}
(\vec{r}_i-\vec{r}_j)^2
\right] + \sum_{i<j<k}^A V_{{\rm NNN}, ijk} \; .
\end{eqnarray}
Next we divide the $A$-nucleon infinite HO basis space
into the finite active space ($P$) comprising of all states of up to $N_{\rm max}$
HO excitations above the unperturbed ground state and the excluded space ($Q=1-P$).  
The basic idea of the NCSM approach is to apply a unitary transformation
on the Hamiltonian (\ref{hamomega}), $e^{-S} H_A^\Omega e^S$ such that
$Q e^{-S} H_A^\Omega e^S P=0$. If such a transformation is found, the effective
Hamiltonian that exactly reproduces a subset of eigenstates of the full space Hamiltonian
is given by $H_{\rm eff}=P e^{-S} H_A^\Omega e^S P$. This effective Hamiltonian
contains up to $A$-body terms and to construct it is essentially as difficult as to solve
the full problem. Therefore, we apply this basic idea on a sub-cluster level.
When a genuine TNI is considered, the simplest approximation is a three-body
effective interaction approximation. The NCSM calculation is then performed with
the following four steps:

(i) We solve a three-nucleon system for all possible three-nucleon channels 
characterized by a fixed angular momentum, isospin and parity 
with the Hamiltonian (\ref{hamomega}), i.e.,
using 
\begin{equation}\label{h3}
{\cal H}_3= h_1+h_2+h_3+V_{12}^{\Omega,A}+V_{13}^{\Omega,A}+V_{23}^{\Omega,A}+V_{{\rm NNN}, 123} \; .
\end{equation}
Note that in Eq. (\ref{h3}), $A>3$ corresponding to the investigated nucleus is kept 
in $V_{ij}^{\Omega,A}, i,j\equiv 123$ as defined in (\ref{hamomega}). Consequently, 
the three nucleons feel a pseudo-mean field of the spectator nucleons generated 
by the HO CM potential. It is necessary to separate the three-body effective interaction 
contributions from the TNI and from the two-nucleon interaction. Therefore, we need to find
three-nucleon solutions for the Hamiltonian with and without the $V_{{\rm NNN}, 123}$ TNI term.
The three-nucleon solutions are obtained by procedures described in Refs. 
\cite{fourb_NCSM,Jacobi_NCSM} (without TNI) and \cite{NCSM_TM} (with TNI).
In particular, we note that we use two-body effective interaction corresponding to
a very large three-nucleon basis space, i.e., up to $N_{\rm max}^{(3)}=40$, to obtain
these solutions. The number of three-nucleon channels to include in the calculations
is restricted by the maximal active space $N_{max}$ to be used in the final $A$-nucleon calculation. 

(ii) We construct the unitary transformation 
corresponding to the choice of the active basis space $P_3$
from the three-nucleon solutions using the Lee-Suzuki procedure \cite{LS1,LS2,LS3,UMOA}. 
In particular, we demand that $Q_3 e^{-S} {\cal H}_3 e^S P_3=0$.
Let us remark that the definition of the three-nucleon active space $P_3$
is determined by the definition of the $A$-nucleon active space $P$. 
The $P_3$ states contains all three-nucleon states up to the highest 
possible three-nucleon excitation, which can be found in the $P$ space
of the $A$-nucleon system. For example,
for $A=6$ and $N_{\rm max}=6$ ($6\hbar\Omega$) space we have $P_3$ defined 
by $N_{\rm 3max}=8$. Similarly, for the $p$-shell nuclei with $A\geq 7$
and $N_{\rm max}=6$ ($6\hbar\Omega$) space we have $N_{\rm 3max}=9$.
We note that the anti-hermitian transformation operator $S$ can be expressed as \cite{UMOA}
\begin{equation}\label{UMOAsol}
S = {\rm arctanh}(\omega-\omega^\dagger)    \; ,
\end{equation}
with the operator $\omega$ satisfying $\omega=Q_3\omega P_3$.
If the eigensystem of the Hamiltonian ${\cal H}_3$ (\ref{h3}) is given by 
${\cal H}_3|k\rangle = E_k |k\rangle $, 
then the operator $\omega$ can be determined from
\begin{equation}\label{omegasol}
\langle\alpha_Q|\omega|\alpha_P\rangle = \sum_{k \in{\cal K}}
\langle\alpha_Q|k\rangle\langle\tilde{k}|\alpha_P\rangle \; ,
\end{equation}  
where the tilde denotes the inverse of the matrix defined by matrix elements  
$\langle\alpha_P|k\rangle$, i.e.,
$\sum_{\alpha_P}\langle\tilde{k}|\alpha_P\rangle\langle\alpha_P
|k'\rangle = \delta_{k,k'}$ and
$\sum_k \langle\alpha'_P|\tilde{k}\rangle \langle k|\alpha_P\rangle 
= \delta_{\alpha'_P,\alpha_P}$, for $k,k'\in{\cal K}$. In Eq.
(\ref{omegasol}),  $|\alpha_P\rangle$ and $|\alpha_Q\rangle$
are the active-space and the $Q$-space basis states, respectively,  
and ${\cal K}$ denotes a set of $d_P$ eigenstates, whose properties
are reproduced in the $P$-space, 
with $d_P$ equal to the dimension of the $P$-space. 
The set ${\cal K}$ is typically chosen by selecting the lowest three-nucleon
eigenstates for the given three-nucleon channel.
In practice, to calculate the $P$-space matrix elements of the effective Hamiltonian,
we apply the Eqs. (15) and (16) of Ref. \cite{C12_NCSM}.  
The three-body effective interaction is then obtained as 
\begin{equation}\label{v3eff}
V_{{\rm 3eff},123}^{\rm NN+NNN}=P_3 \left[ e^{-S_{\rm NN+NNN}}(h_1+h_2+h_3+V_{12}^{\Omega,A}
+V_{13}^{\Omega,A}+V_{23}^{\Omega,A}+V_{{\rm NNN}, 123})e^{S_{\rm NN+NNN}}
-(h_1+h_2+h_3)\right] P_3 \;  
\end{equation}
and
\begin{equation}\label{v3eff_2b}
V_{{\rm 3eff},123}^{\rm NN}=P_3 \left[e^{-S_{\rm NN}}(h_1+h_2+h_3+V_{12}^{\Omega,A}
+V_{13}^{\Omega,A}+V_{23}^{\Omega,A})e^{S_{\rm NN}}
-(h_1+h_2+h_3)\right] P_3 \; . 
\end{equation}
Note that we distinguish the transformation $S_{\rm NN+NNN}$ corresponding to the Hamiltonian
${\cal H}_3$ (\ref{h3}) and the transformation $S_{\rm NN}$ corresponding to the Hamiltonian
without the TNI term, i.e.,
$h_1+h_2+h_3+V_{12}^{\Omega,A}+V_{13}^{\Omega,A}+V_{23}^{\Omega,A}$.
The three-body effective interaction contribution
from the TNI we then define as 
\begin{equation}\label{v3eff_3b}
V_{{\rm 3eff},123}^{\rm NNN}\equiv V_{{\rm 3eff},123}^{\rm NN+NNN}
-V_{{\rm 3eff},123}^{\rm NN} \; .
\end{equation}

(iii) As the three-body effective interactions are derived in the Jacobi-coordinate
HO basis but the $p$-shell nuclei calculations will be performed more efficiently
in a Cartesian-coordinate
single-particle Slater-determinant $M$-scheme basis, we need to perform a suitable 
transformation of the interactions. This transformation is a generalization  
of the well-known transformation on the two-body level that depend on HO 
Brody-Moshinsky brackets. The antisymmetrized Jacobi-coordinate HO basis
in which the three-body effective interaction is obtained can be represented in the form
$|N i J M T M_T\rangle$ where $N$ is the total number of HO excitation, $i$ 
is an additional quantum number enumerating the antisymmetrized states,
$J$ and $T$ are the total three-nucleon angular momentum and isospin, respectively,
and $M$, $M_T$ their third components. This basis can be expanded in a $2+1$ nucleon cluster
basis using expansion coefficients, i.e.,
\begin{equation}\label{Ja_bas}
|N i J M T M_T\rangle = \sum \langle (nlsjt,{\cal NL}\frac{1}{2}{\cal J}\frac{1}{2})||NiJT\rangle
|(nlsjt,{\cal NL}\frac{1}{2}{\cal J}\frac{1}{2})J M T M_T\rangle \; ,
\end{equation}
where the $|nl\rangle$ HO state describes relative motion of nucleon 1 and 2, $|{\cal NL}\rangle$
describes the relative motion of the third nucleon with respect to the CM of nucleons 1 and 2,
and $\langle (nlsjt,{\cal NL}\frac{1}{2}{\cal J}\frac{1}{2})||NiJT\rangle$ is the coefficient
of fractional parentage.
It holds that $N=2n+l+2{\cal N}+{\cal L}$. For further details and the procedure how to calculate
the expansion coefficients see Ref. \cite{Jacobi_NCSM}. The basis for the three-body effective 
interaction suitable for the shell-model code input is a three-nucleon Slater-determinant
HO basis characterized by $M_3=m_{j_1}+m_{j_2}+m_{j_3}$, 
$M_{\rm T}=m_{\rm t_1}+m_{\rm t_2}+m_{\rm t_3}$ and parity,
formed by the single particle states $|nl\frac{1}{2}j m_j\frac{1}{2}m_t\rangle$ 
that depend on the single-particle coordinates. 
We introduce a short-hand notation for these Slater-determinant states, 
i.e., $|(nl\frac{1}{2} j m_j \frac{1}{2} m_t)_{(abc)}\rangle$, where $a,b,c$ labels the 
occupied states. 
In order to transform the three-body effective interaction to the new basis we need first
to couple the relative coordinate basis (\ref{Ja_bas}) with the three-nucleon CM HO states
$|N_{\rm CM} L_{\rm CM} M_{\rm CM}\rangle$ to form a complete basis. The three-body 
effective interaction is independent on the $N_{\rm CM} L_{\rm CM} M_{\rm CM}$
quantum numbers. The overlap of the two respective states can then be expressed 
in the form
\begin{eqnarray}\label{v3trans}
&&\langle (nl\frac{1}{2} jm_j \frac{1}{2} m_t)_{(abc)} | N i J M T M_T ;
N_{\rm CM} L_{\rm CM} M_{\rm CM}\rangle =
\delta_{2n_a+l_a+2n_b+l_b+2n_c+l_c,N+2N_{\rm CM}+L_{\rm CM}}
\nonumber \\
&\times&
\delta_{m_{j_a}+m_{j_b}+m_{j_c},M+M_{\rm CM}}
\delta_{m_{t_a}+m_{t_b}+m_{t_c},M_T}
\sqrt{6} \sum \;
\langle (nlsjt,{\cal NL}\frac{1}{2}{\cal J}\frac{1}{2})||NiJT\rangle
\; \frac{1}{2} (1-(-1)^{l+s+t})
\nonumber \\
&\times&
(l_a m_a \frac{1}{2} m_{s_a} | j_a m_{j_a})
(l_b m_b \frac{1}{2} m_{s_b} | j_b m_{j_b})
(l_c m_c \frac{1}{2} m_{s_c} | j_c m_{j_c})
(\frac{1}{2} m_{t_a} \frac{1}{2} m_{t_b} | t m_t)
(t m_t \frac{1}{2} m_{t_c} | T M_T)
\nonumber \\
&\times&
(\frac{1}{2} m_{s_a} \frac{1}{2} m_{s_b} | s m_s)
(l_b m_b l_a m_a | \Lambda m_{\Lambda})
(L_{12} M_{12} l m_l | \Lambda m_{\Lambda})
(l_c m_c L_{12} M_{12} | \lambda m_{\lambda})
\nonumber \\
&\times&
(L_{\rm CM} M_{\rm CM} {\cal L} {\cal M}_{\cal L}|\lambda m_{\lambda})
({\cal L} {\cal M}_{\cal L} \frac{1}{2} m_{s_c}|
{\cal J} {\cal M}_{\cal J})
(l m_l s m_s|j m_j)
(j m_j {\cal J} {\cal M}_{\cal J}| J M)
\nonumber \\
&\times&
\langle n_c l_c N_{12} L_{12} \lambda | N_{\rm CM} L_{\rm CM}
{\cal N L} \lambda \rangle_{\frac{1}{2}}
\langle n_b l_b n_a l_a \Lambda | N_{12} L_{12} n l \Lambda \rangle_{1} \; ,
\end{eqnarray}
where we explicitly show the total HO quantum number conservation that also implies
parity conservation, the angular momentum and isospin third component conservation.
The sum goes over the expansion (\ref{Ja_bas}), $\lambda$, $\Lambda$, $N_{12}$, $L_{12}$ 
and the $m$ quantum numbers in the  Clebsch-Gordan coefficients not appearing 
on the left-hand side. The 
$\langle n_1 l_1 n_2 l_2 L | n_3 l_3 n_4 l_4 L \rangle_{d}$ are the generalized
Brody-Moshinsky brackets for two-particles of mass ratio $d$ as defined e.g., in 
Ref. \cite{Trlifaj}.
Sums of the Clebsch-Gordan coefficients in Eq. (\ref{v3trans}) can be re-expressed as 
sums of 6-j and 9-j coefficients. However, some Clebsch-Gordan coefficients will
remain as we have a magnetic quantum number dependence on the left-hand side.
It should be pointed out that the overlap (\ref{v3trans}) is independent of the HO frequency
$\Omega$, and for a given $N_{\rm max}$, it needs to be calculated just once.
The three-body effective interaction $M$-scheme matrix elements are then obtained by
\begin{eqnarray}\label{v3trans_fin}
\langle (nl\frac{1}{2} jm_j \frac{1}{2} m_t)_{(abc)} |V_{\rm 3eff,123}|
(nl\frac{1}{2} jm_j \frac{1}{2} m_t)_{(def)} \rangle \; &&= \sum \;
\langle (nl\frac{1}{2} jm_j \frac{1}{2} m_t)_{(abc)} |
N i J M T M_T ; N_{\rm CM} L_{\rm CM} M_{\rm CM}\rangle
\nonumber \\ 
\langle N i J T |V_{\rm 3eff,123}|N' i' J T\rangle \;
&&
\langle N' i' J M T M_T ; N_{\rm CM} L_{\rm CM} M_{\rm CM}|
(nl\frac{1}{2} jm_j \frac{1}{2} m_t)_{(def)} \rangle \; .
\end{eqnarray}

(iv) We solve the Schr\"odinger equation for the $A$-nucleon system using the Hamiltonian 
\begin{equation}\label{Ham_A_Omega_eff}
H^\Omega_{A, {\rm eff}}=\sum_{i=1}^A h_i 
+ \frac{1}{A-2}\sum_{i<j<k}^A V_{{\rm 3eff}, ijk}^{\rm NN}
+\sum_{i<j<k}^A V_{{\rm 3eff}, ijk}^{\rm NNN} \; ,
\end{equation}
where the $\frac{1}{A-2}$ factor takes
care of over-counting the contribution from the two-nucleon interaction. At this point
we also subtract the $H_{\rm CM}$ and add the Lawson projection term 
$\beta(H_{\rm CM}-\frac{3}{2}\hbar\Omega)$ to shift the spurious
CM excitations. Just as the bare interaction is translationally invariant, so is
the effective interaction, and the energies of the physical eigenstates
corresponding to the $0\hbar\Omega$ excitation of the CM are independent 
on the choice of $\beta$. 
The $A$-nucleon calculation is then
performed using either the Many-Fermion Dynamics shell model code \cite{MFD} generalized 
to handle three-body interactions or using the newly developed code REDSTICK \cite{OrmJohn}, using 
a different algorithm for evaluating the Hamiltonian matrix elements, 
with the Hamiltonian in the form: 
\begin{equation}\label{Ham_A_Omega_eff_SM}
H^{\Omega, {\rm SM}}_{A, {\rm eff}}=\sum_{i<j}^A \left[ \frac{(\vec{p}_i-\vec{p}_j)^2}{2Am}
+\frac{m\Omega^2}{2A} (\vec{r}_i-\vec{r}_j)^2\right]
+ \frac{1}{A-2}\sum_{i<j<k}^A V_{{\rm 3eff}, ijk}^{\rm NN}
+\sum_{i<j<k}^A V_{{\rm 3eff}, ijk}^{\rm NNN} + \beta(H_{\rm CM}-\frac{3}{2}\hbar\Omega)\; .
\end{equation}

We summarize this section by repeating that the effective interaction depends on the nucleon
number $A$, the HO frequency $\Omega$, and the $P$-space basis size defined by 
$N_{\rm max}$. It is 
translationally invariant and with $N_{\rm max}\rightarrow \infty$ the NCSM effective
Hamiltonian approaches the starting bare Hamiltonian (\ref{ham}). Consequently, with the basis
size increase the NCSM results become less and less dependent on the HO frequency
and converge to the exact solution. Alternatively, by increasing the clustering
of the effective interaction for a fixed $P$-space size the NCSM effective Hamiltonian
approaches the exact $A$-nucleon effective Hamiltonian that reproduces exactly a subset of
eigenstates of the starting Hamiltonian (\ref{ham}).

\section{Application to selected $p$-shell nuclei}\label{sec_res}

The calculations with the three-body interactions are technically and computationally
intensive. When a genuine TNI is employed the step (i) from the previous section is 
computationally demanding. In addition, when a three-body effective interaction is used
in general, the steps (iii) and (iv) are complex as well.
In these first NCSM calculations with the TNI in the $p$-shell, we limit ourselves
to the use of a single realistic interaction. In particular, we use the isospin invariant
AV8$^\prime$ NN potential \cite{av8p} that is a simplified version of the AV18 NN potential.
For the TNI, we employ a version of the original Tucson-Melbourne 
force \cite{TM} corrected to drop a term later found to be inconsistent with the
chiral-symmetry-based effective field theory TNI of two-pion range \cite{Bira}.  
We note that there have been different strength constants (taken from
pion-nucleon scattering data) suggested for this force \cite{TM,TMprime99,Bira}.  
We use only one particular set, i.e. the TM$^\prime$(99) updated  strength constants of the
Tucson-Melbourne TNI that was introduced in Ref. \cite{TMprime99}.  
We summarize the strength constants and other numbers employed in the present
investigation in Table \ref{TMparam}.  The cutoff $\Lambda$ is the only parameter 
in the TNI and it is set to $\Lambda = 4.7$ by fitting the $^3$H binding energy using
the AV8$^\prime$ + TM$^\prime$(99) interactions.
We add the Coulomb potential to the Hamiltonian perturbatively,
that is only in the $P$-space.

The NCSM calculations depend in general on the HO frequency and the basis size.
The basis size is limited by the available computational resources and the algorithm
development. In this paper, we were able to reach the $6\hbar\Omega$ basis space for
$A=6,7$. For higher masses, we limited ourselves to the $4\hbar\Omega$ basis space.
The optimal HO frequency is selected in the NCSM by the requirement of the least
dependence of the eigenergies on the frequency. Typically, for the $p$-shell nuclei
we select the optimal frequency by finding the ground-state energy minimum
in the largest accessible basis space. Due to the complexity of the calculations
with the TNI, in this study we rely on our previous investigations done without the
TNI to select the optimal frequency and perform comparative calculations with and without
the TNI. More detailed frequency dependence investigations will be done later.

The steps (i) and (ii) of the calculations as described in Section \ref{sec_theory}
are performed using the Jacobi-coordinate many-body effective interaction code ({\it manyeff})
\cite{Jacobi_NCSM} extended to handle the TNI as described in Ref. \cite{NCSM_TM}.
When calculating the three-body effective interaction induced by the two-nucleon
potential, $V_{{\rm 3eff},123}^{\rm NN}$ (\ref{v3eff_2b}), we are able reach large
three-nucleon basis spaces, e.g. $N^{(3)}_{\rm max}=40$, to obtain high precision
three-nucleon solutions. Due to the complexity of the calculations with the TNI
we were, however, limited in this work to smaller spaces, up to $N^{(3)}_{\rm max}=30$ 
\cite{NCSM_TM}, when the TNI was included in the Hamiltonian (\ref{h3}). 
This was sufficient to obtain the $^3$H binding energy within 50 keV of the exact
result, see Fig. 2 of Ref. \cite{NCSM_TM}. 
Nevertheless, it would still be useful to further increase $N^{(3)}_{\rm max}$ 
in the future. In particular because here we employ a lower HO frequency 
for which we expect a slower convergence than for that used
in the $^3$H calculations. In order
to calculate the $V_{{\rm 3eff},123}^{\rm NNN}$ (\ref{v3eff_3b}) it is essential
to use the same $N^{(3)}_{\rm max}$ to obtain (\ref{v3eff}) and (\ref{v3eff_2b}).
Therefore, we performed three calculations, one up to $N^{(3)}_{\rm max}=40$
to calculate $V_{{\rm 3eff},123}^{\rm NN}$ (\ref{v3eff_2b}) in all relevant
three-nucleon channels and two for (\ref{v3eff}) and (\ref{v3eff_2b})
in the same basis space up to $N^{(3)}_{\rm max}=30$ 
for the three-nucleon channels up to $J=\frac{7}{2}$ 
to calculate $V_{{\rm 3eff},123}^{\rm NNN}$ (\ref{v3eff_3b}).
We will improve on this in our future studies as also discussed in the concluding section.   
The step (iii) is performed by a specialized code that reads the outputs
of the {\it manyeff} code and the step (iv) is then
performed using either the Many-Fermion Dynamics shell model code \cite{MFD} generalized 
to handle three-body interactions or using the newly developed code REDSTICK \cite{OrmJohn}. 
 
In order to get the most accurate assessment of the effect of the genuine three-nucleon interaction
in the following we always compare calculation with and without the TNI using identical 
basis size, HO frequency and the clustering approximation, i.e., the three-body effective 
interaction.

We note that bare nucleon charges and unrenormalized transition operators were used 
in all the transition calculations presented in this paper. For some transitions,
E2 in particular, the operator renormalization could be important. The operator
renormalization can be calculated as outlined, e.g. in Ref. \cite{C12_NCSM},
and is now under development. In the Gamow-Teller transition
calculations, the $g_A$ constant was not included in the presented B(GT) values. 

Finally, let us remark that all the presented calculated states in this paper are the $p$-shell
or the $0\hbar\Omega$ dominated states. The opposite parity ($1\hbar\Omega$) and the
$2\hbar\Omega$-dominated states were investigated within the NCSM, e.g. in Refs. 
\cite{C12_NCSM,v3eff,Be8_NCSM,A10_NCSM}. Such states are not obtained at low enough energy
in the basis spaces used in the present investigation. Therefore, in this paper comparison
is made with experimental levels known to be predominantly of $p$-shell character.

\subsection{$^6$Li, $^6$He}\label{subs_li6}

\begin{figure}
\vspace*{2cm}
\includegraphics[width=7.0in]{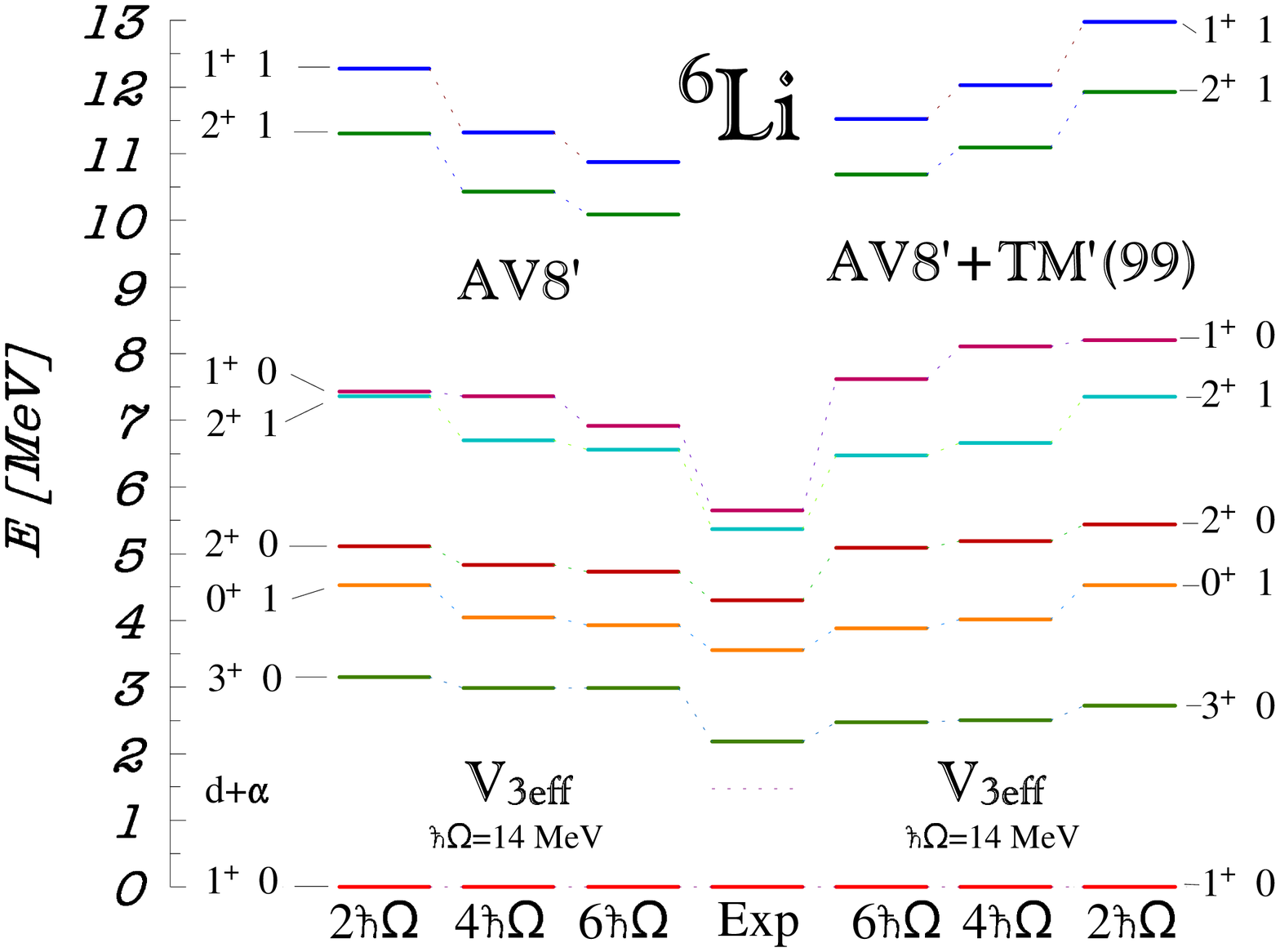}
\caption{\label{li6_exc_TM} Calculated positive-parity excitation spectra of
$^{6}$Li obtained in $2\hbar\Omega$-$6\hbar\Omega$ basis spaces using three-body effective
interactions derived from AV8$^\prime$ NN potential and AV8$^\prime$ NN potential
plus TM$^\prime$(99) three-nucleon interaction, respectively,
are compared to experiment. The HO frequency of $\hbar\Omega=14$ MeV was used.
The experimental values are from Ref. \protect\cite{AS88}.
}
\end{figure}

We investigated $A=6$ nuclei using two-nucleon interactions in Ref. \cite{NCSM_6} and
$^6$Li in particular in Ref. \cite{v3eff} where we used three-body effective
interaction derived from the AV8$^\prime$.
Our $^6$Li results with the TNI are summarized in Table \ref{tab_li6} and Fig. \ref{li6_exc_TM}.  
We show the basis size dependence up to $6\hbar\Omega$ using the optimal frequency
$\hbar\Omega=14$ MeV. It is apparent from Fig. \ref{li6_exc_TM} that the basis 
size dependence is quite similar
in the calculations with and without the TNI and it is important that the excitation
energies of the low-lying states in particular show a good stability and a convergence 
pattern when the $6\hbar\Omega$ basis space is reached. At the same time we can see
a significant sensitivity of the excitation spectra to the presence of the genuine 
three-nucleon interaction. It should be noted that the $3^+_1 0$ state excitation
energy decreases significantly and it gets closer to experiment when the TNI is included.
Also, the $3^+_1 0 \leftrightarrow 2^+_1 0$ splitting, which is  underestimated with just
a two-nucleon interaction, increases significantly. Our calculated $^6$Li binding 
energy increases by more than 2.5 MeV when the TNI is included. With tighter binding with 
the TNI the point-proton radius decreases slightly similarly as the B(E2) value of 
the $1^+_1 0\rightarrow 3^+_1 0$ transition. The ground-state quadrupole moment and the M1 
transitions are not much affected, although there appears to be an improvement
for the $2^+_1 1\rightarrow 1^+_1 0$ M1 transition.  
 
Our $^6$He results are summarized also in Table \ref{tab_li6}. Apart from increased binding energy,
there are little differences between the calculations with and without the TNI.
In the same table, we show the B(GT) values obtained for the $^6$He$\rightarrow ^6$Li
ground-state to ground-state transition. Clearly, the TNI shifts the result in the right 
direction closer to experiment although a discrepancy still remains. Recently, Schiavilla 
and Wiringa found that the AV18/Urbana-IX interaction also over-predicts the 
$^6$He$\rightarrow ^6$Li B(GT) value \cite{Schia}. In fact, their obtained Gamow-Teller
matrix elements ($\equiv\sqrt{{\rm B(GT)}}$), 2.254(5) and 2.246(10) using two types of wave functions,
compare well with our result 2.283 for AV8$^\prime$+TM$^\prime$(99). Our corresponding
AV8$^\prime$ result is 2.305 which is basically identical to our CD-Bonn result presented 
in Ref. \cite{NCSM_6}.

\subsection{$^7$Li, $^7$Be}\label{subs_li7}

\begin{figure}
\vspace*{2cm}
\includegraphics[width=7.0in]{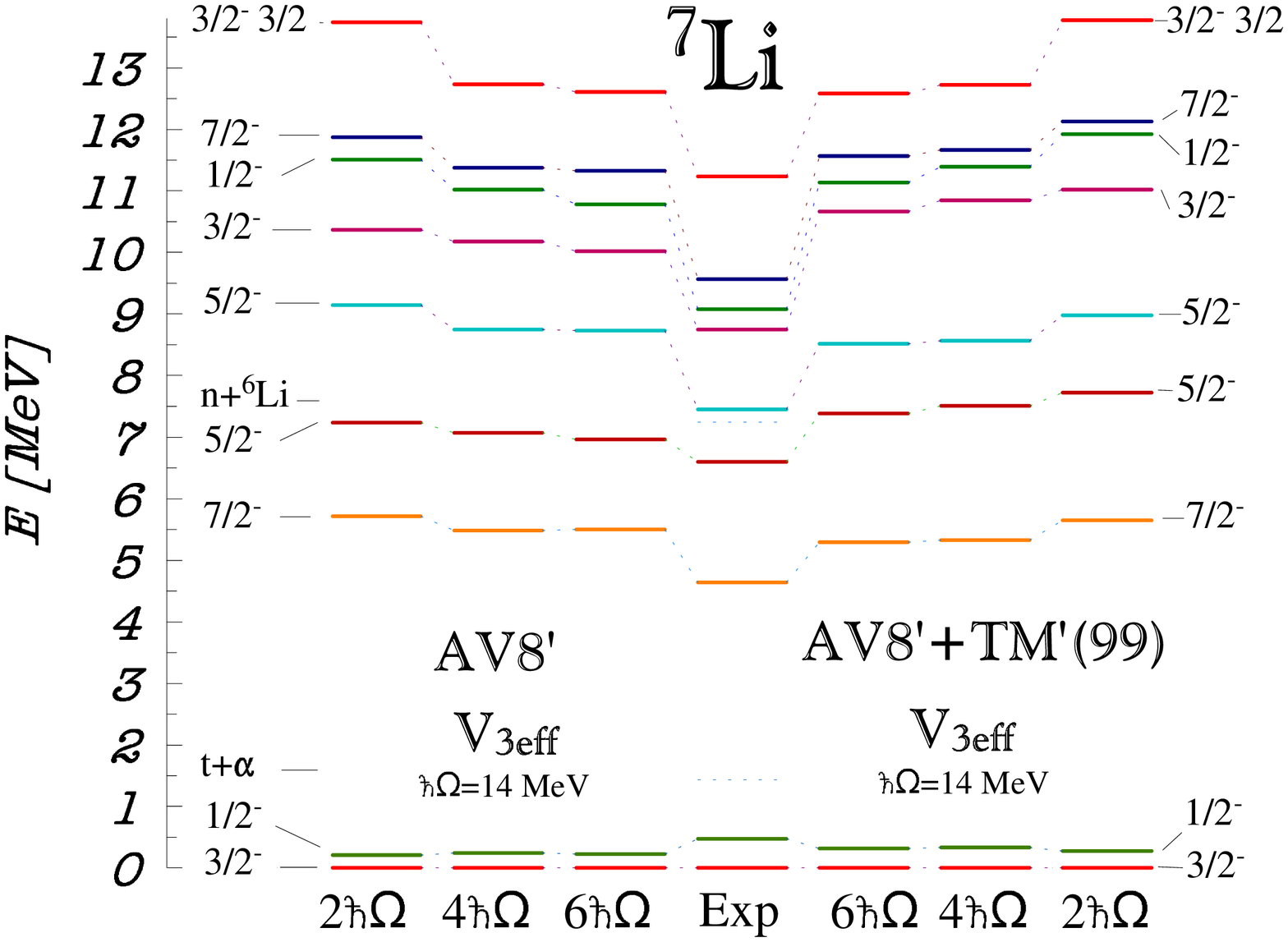}
\caption{\label{li7_exc_TM} Calculated negative-parity excitation spectra of
$^{7}$Li obtained in $2\hbar\Omega$-$6\hbar\Omega$ basis spaces using three-body effective
interactions derived from AV8$^\prime$ NN potential and AV8$^\prime$ NN potential
plus TM$^\prime$(99) three-nucleon interaction, respectively,
are compared to experiment. The HO frequency of $\hbar\Omega=14$ MeV was used.
The experimental values are from Ref. \protect\cite{Till02}.
}
\end{figure}

Our $^7$Li results with the TNI are summarized in Table \ref{tab_li7} and Fig. \ref{li7_exc_TM}.  
We show the basis size dependence up to $6\hbar\Omega$ using the optimal frequency
$\hbar\Omega=14$ MeV. We note that the $6\hbar\Omega$ $^7$Li calculation is the most
complex one among those presented in this paper. The three-body effective interaction
basis space is defined by $N_{\rm 3max}=9$ and there are 741 823 056 three-nucleon 
$M$-scheme matrix elements that we input in the shell model code. 
Although the dimension of the Hamiltonian
matrix is only  663 527, it is significantly less sparse when a three-body interaction
is present, e.g., there are 4 555 058 857 non-zero matrix elements  
compared to 235 927 305 for a calculations with just a two-body interaction. 

From our $^7$Li results we can draw similar conclusions as from the $^6$Li results.
A particularly important observation is the excitation energy convergence pattern,
i.e., good stability of the spectra when increasing the basis size from $4\hbar\Omega$ 
to $6\hbar\Omega$ for both calculations with the TNI and without the TNI. There is
an expected increase of binding energy and sensitivity of the spectra to the presence
of the TNI. The level ordering is unchanged for the 9 levels that we show. 
Note that when compared to the old experimental evaluation of Ref. \cite{AS88}
we obtain the
reversed order of $\frac{3}{2}^-_2 \frac{1}{2}$ and $\frac{7}{2}^-_2 \frac{1}{2}$ states,
consistently with previous NCSM studies, e.g., \cite{NB98}. However,
the new evaluation \cite{Till02} changes the order and energies of the two levels
and, in addition, adds a new $\frac{1}{2}^- \frac{1}{2}$ level. Our calculated level ordering 
then matches the new evaluated level ordering. Note that our calculation with the TNI
shows a slightly improved relative level spacing for those three levels compared to experiment.  
Similarly, the relative level spacing of the $\frac{7}{2}^-_1 \frac{1}{2}$, 
$\frac{5}{2}^-_1 \frac{1}{2}$ and $\frac{5}{2}^-_2 \frac{1}{2}$ states is improved compared to 
experiment in the calculation with the TNI. Also we note that the lowest two
excited states, $\frac{1}{2}^-_1 \frac{1}{2}$ and $\frac{7}{2}^-_1 \frac{1}{2}$, 
move closer to experiment when the TNI is present.

In Table \ref{tab_li7}, we also show our results for $^7$Be and, in particular,
the Gamow-Teller transitions from the $^7$Be ground state to the ground state
and the first excited state of $^7$Li. There is a slight overall improvement 
when the TNI is included. The same transitions were calculated by Schiavilla and Wiringa
using the the AV18/Urbana-IX interaction \cite{Schia}. Our Gamow-Teller matrix element
results compare well with theirs in particular for the ground-state to ground-state transition.

\subsection{$^8$Be}\label{subs_be8}

\begin{figure}
\vspace*{2cm}
\includegraphics[width=8.0in]{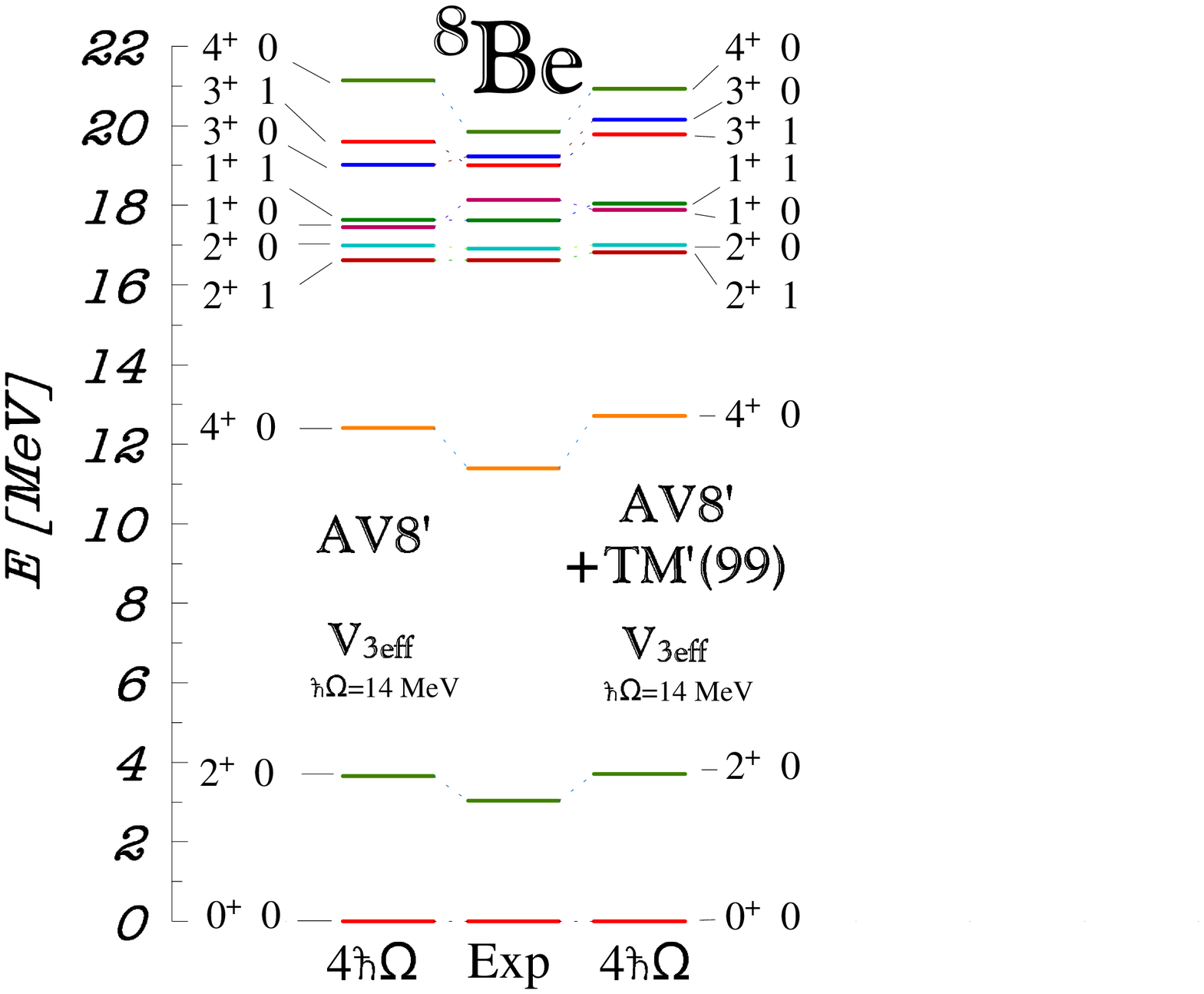}
\caption{\label{be8_exc_TM} Calculated positive-parity excitation spectra of
$^{8}$Be obtained in $4\hbar\Omega$ basis space using three-body effective
interaction derived from AV8$^\prime$ NN potential and AV8$^\prime$ NN potential
plus TM$^\prime$(99) three-nucleon interaction, respectively,
are compared to experiment. The HO frequency of $\hbar\Omega=14$ MeV was used.
The experimental values are from Ref. \protect\cite{AS88}.
}
\end{figure}

We investigated $^8$Be with an emphasis on the intruder states 
using two-nucleon interactions in Ref. \cite{Be8_NCSM} and also in Ref. \cite{v3eff}.
Our $^8$Be results with the TNI are summarized in Table \ref{tab_be8} and Fig. \ref{be8_exc_TM}.  
We compare results obtained using three-body effective interaction derived
from the AV8$^\prime$ two-nucleon interaction and from the AV8$^\prime$+TM$^\prime$(99)
interaction, respectively, in the $4\hbar\Omega$ basis space using the optimal frequency
$\hbar\Omega=14$ MeV. It should be pointed out that our excitation spectra calculation 
without the TNI is actually in a very reasonable agreement with experiment. It is
comforting that turning on the TM$^\prime$(99) three-nucleon interaction does not really
change the excitation spectrum very much and the quality of agreement with experiment
is about the same. With the TNI we gain almost 4 MeV of binding energy.   
Also, we observe a stronger isospin mixing of the $2^+$ and the $1^+$ $T=0,1$ doublets.
We note that in experiment, the $2^+$ states at 16.6 and 16.9 MeV have strongly
mixed $T=0$ and $T=1$ components. In Table \ref{tab_be8}, we present the
intruder $0^+_2 0$ state that was subject of our study in Ref. \cite{Be8_NCSM}.
The position of this state, which is dominated by higher than $0\hbar\Omega$ components
in the wave function, is strongly dependent on the basis size and still a much larger
basis is needed for convergence. In Ref. \cite{Be8_NCSM}, we investigated this 
state using basis spaces up to $10\hbar\Omega$, although employing just the two-body 
effective interaction. We note that the TNI affects this state by increasing its energy
from 26.5 MeV by about 1 MeV in the $4\hbar\Omega$ space. However, this is not significant
as we extrapolate the position of this state with the basis size increase to be in the
10-15 MeV range \cite{Be8_NCSM}.

\subsection{$^{10}$B, $^{10}$Be, $^{10}$C}\label{subs_b10}

\begin{figure}
\vspace*{2cm}
\includegraphics[width=8.0in]{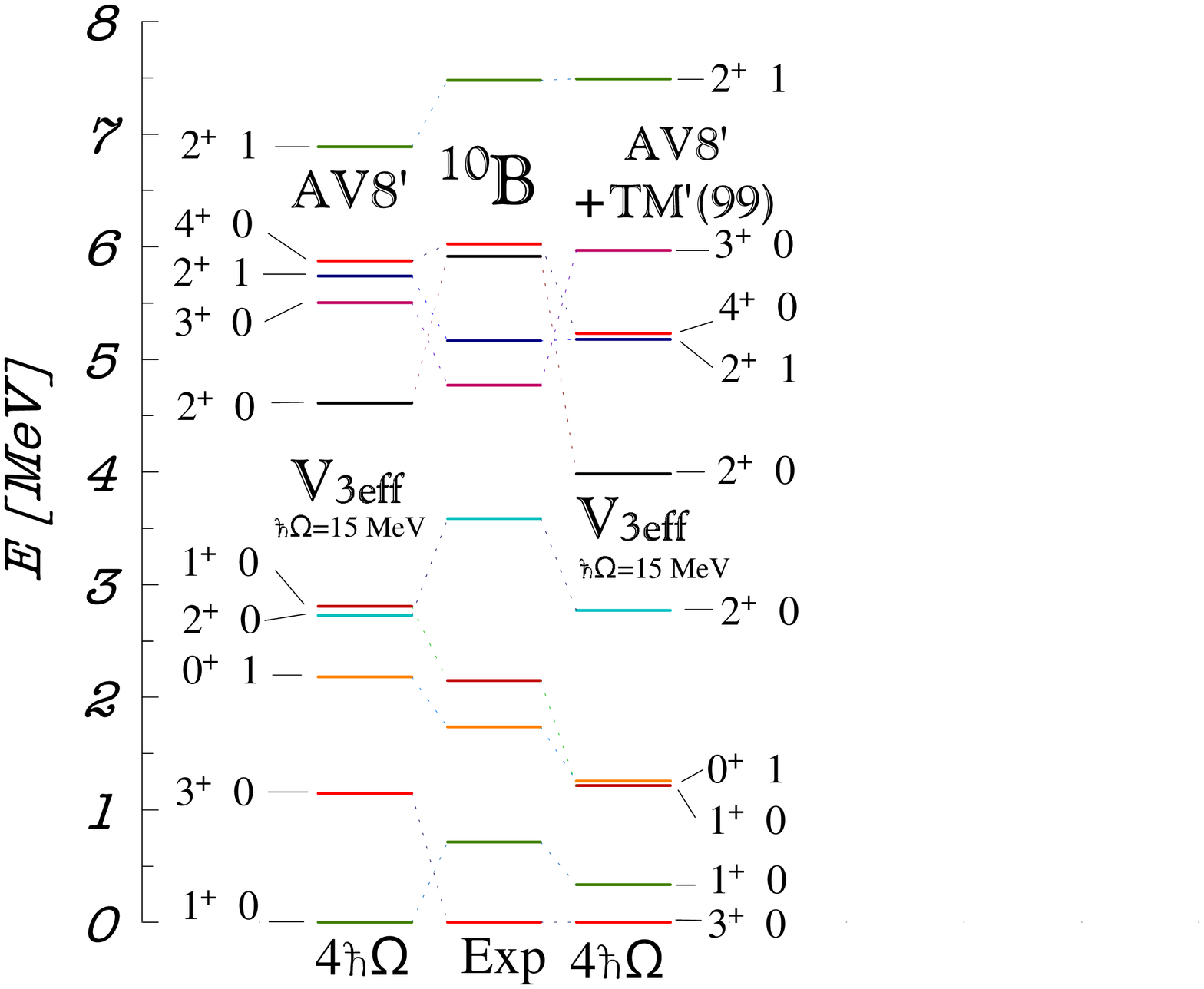}
\caption{\label{b10_exc_TM} Calculated positive-parity excitation spectra of
$^{10}$B obtained in $4\hbar\Omega$ basis space using three-body effective
interaction derived from AV8$^\prime$ NN potential and AV8$^\prime$ NN potential
plus TM$^\prime$(99) three-nucleon interaction, respectively,
are compared to experiment. The HO frequency of $\hbar\Omega=15$ MeV was used.
The experimental values are from Ref. \protect\cite{AS88}.
}
\end{figure}

Detailed $A=10$ NCSM calculations using realistic two-nucleon interactions
were presented in Ref. \cite{A10_NCSM}. In addition, we investigated
$^{10}$B using three-body effective interaction derived from the AV8$^\prime$
NN interaction in Ref. \cite{v3eff}. Our $^{10}$B results obtained with the TNI
are compared to the two-nucleon interaction calculation in Table \ref{tab_b10} 
and Fig. \ref{b10_exc_TM}. We present only experimental states known to be predominantly 
of $p$-shell character as discussed earlier. In $^{10}$B there are $sd$-shell dominated 
low-lying states, $1^+ 0$ at 5.18 MeV and $0^+ 1$ at 7.56 MeV that we obtain at much higher 
excitation energies in the basis spaces employed in this investigation \cite{A10_NCSM}.    
Note that in Fig. \ref{b10_exc_TM}, the two-nucleon
interaction excitation energy results are displayed relative to the $1^+_1 0$ state
while in Table \ref{tab_b10} the same results are given relative to the $3^+_1 0$
state. We note that our calculations with realistic two-nucleon 
interactions predict an incorrect ground-state spin in this nucleus~\cite{v3eff,A10_NCSM}, 
i.e., $1^+_1 0$ instead
of the experimental $3^+_1 0$. The same prediction is obtained by the GFMC method 
\cite{GFMC_9_10}. Therefore, it is particularly interesting and important to investigate
this nucleus using a Hamiltonian that includes a realistic three-nucleon interaction.
Our results obtained using the TM$^\prime$(99) TNI indeed show that 
the three-nucleon interaction has a positive impact on the excitation spectrum and
corrects the level ordering by bringing the $3^+_1 0$ state below the $1^+_1 0$
state in agreement with experiment. Level ordering improves also for other excited
states. This is particularly apparent when we present the excitation energies
relative to the $3^+_1 0$ state as done in Table \ref{tab_b10}. In the GFMC calculations
of Ref. \cite{GFMC_9_10}, the correct $^{10}$B ground state was obtained using
the Illinois TNI but the $1^+_1 0$ ground state remained when the Urbana IX TNI
was employed. Combined with our observed sensitivity for higher excited states 
this shows that nuclear structure will play an important role in determining
the form and the parametrization of the three-nucleon interaction that, unlike 
the two-nucleon interaction, is still not well known.
Concerning the binding energies, with the TM$^\prime$(99) we gain almost 6 MeV for the
$3^+_1 0$ state and about 4 MeV for the $1^+_1 0$ state. 

Our $^{10}$Be and $^{10}$C results are presented also in Table \ref{tab_b10}.
In particular, we paid attention to the Gamow-Teller transitions from the
$^{10}$B ground state to the $2^+ 1$ states of $^{10}$Be and the transition
from the $^{10}$C ground state to the $1^+_1 0$ state of $^{10}$B. Overall,
we observe a clear improvement compared to experiment in calculations
that include the TNI. In the excitation spectra, we note the increased
splitting between the $2^+_1 1$ and the $2^+_2 1$ states when TNI is included,
again an improvement compared to experiment.

\subsection{$^{11}$B, $^{11}$C}\label{subs_b11}

\begin{figure}
\vspace*{2cm}
\includegraphics[width=8.0in]{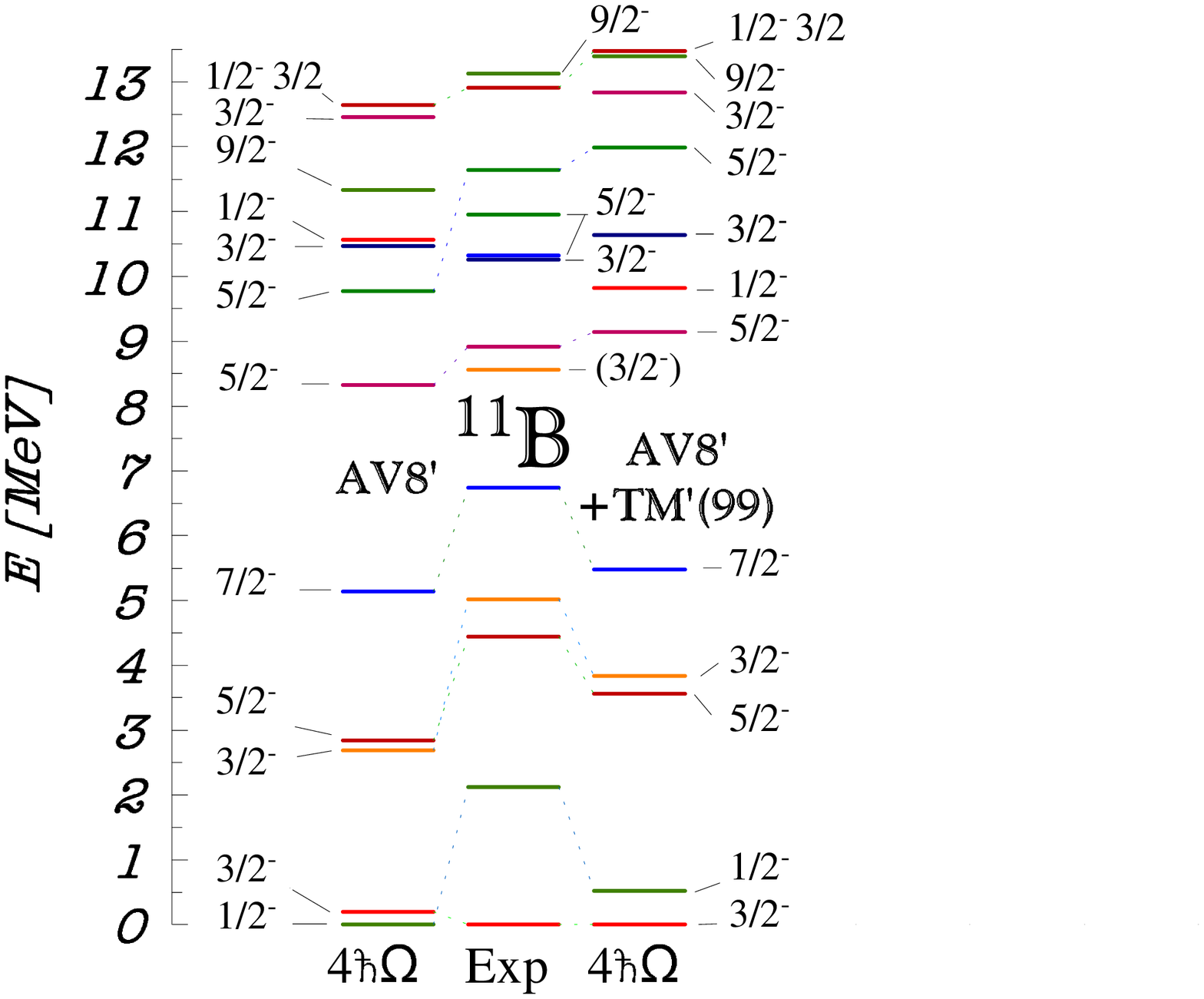}
\caption{\label{b11_exc_TM} Calculated negative-parity excitation spectra of
$^{11}$B obtained in $4\hbar\Omega$ basis space using three-body effective
interaction derived from AV8$^\prime$ NN potential and AV8$^\prime$ NN potential
plus TM$^\prime$(99) three-nucleon interaction, respectively,
are compared to experiment. The HO frequency of $\hbar\Omega=15$ MeV was used.
Assignments are made only to experimental states that are known to be dominantly 
of $p$-shell character.
The experimental values are from Ref. \protect\cite{AS90,Arya85}.
}
\end{figure}

There were no published NCSM calculations for $^{11}$B up to now.
Our $^{11}$B results with and without the TNI are summarized in Table \ref{tab_b11} 
and Fig. \ref{b11_exc_TM}. We compare results obtained using three-body effective 
interaction derived
from the AV8$^\prime$ two-nucleon interaction and from the AV8$^\prime$+TM$^\prime$(99)
interaction, respectively, in the $4\hbar\Omega$ basis space using the same optimal frequency
as for $^{10}$B, i.e., $\hbar\Omega=15$ MeV. All shown calculated states are $0\hbar\Omega$
dominated. It is not straightforward to make a correct correspondence to the experimental
$T=\frac{1}{2}$ levels in particular above 10 MeV of excitation energy. We follow 
Refs. \cite{Wolt90,Millener_pr} in making assignments in Table \ref{tab_b11} 
and Fig. \ref{b11_exc_TM}.
The $T=\frac{3}{2}$ level energies are taken from Refs. \cite{Arya85,Millener}.  
Similarly as for $^{10}$B, in Fig. \ref{b11_exc_TM}, the two-nucleon
interaction excitation energy results are displayed relative to 
the $\frac{1}{2}^-_1 \frac{1}{2}$ state
while in Table \ref{tab_b11} the same results are given relative to 
the $\frac{3}{2}^-_1 \frac{1}{2}$ state.  
Given the complicated situation in neighboring $^{10}$B, it is quite interesting to 
investigate $^{11}$B. Although we have not done a detailed frequency dependence study
as for $^{10}$B, our present calculation suggest that the AV8$^\prime$ NN potential
by itself might produce the incorrect ground state $\frac{1}{2}^-_1 \frac{1}{2}$
instead of the experimentally observed $\frac{3}{2}^-_1 \frac{1}{2}$ state. 
More calculations are needed
to explore this issue but based on our present calculations we can definitely expect 
that a realistic two-nucleon interaction will predict the $\frac{3}{2}^-_1$ and
$\frac{1}{2}^-_1$ states almost degenerate. In our $4\hbar\Omega$ calculation
with the AV8$^\prime$ we obtain an incorrect ground state spin for $^{11}$B.
By including the TM$^\prime$(99) this is corrected together with the level
ordering of $\frac{5}{2}^-_1 \frac{1}{2}$ and $\frac{3}{2}^-_2 \frac{1}{2}$
states. In addition, the level spacing improves for majority of the low-lying
states shown in Fig. \ref{b11_exc_TM} and Table \ref{tab_b11}. The binding energy
gain is about 6 MeV.  

As for $^{11}$B, the calculation with the TNI produces a superior excitation
spectrum and the correct ground-state spin for $^{11}$C. An exception where there is 
a shift away from experiment in the calculations with the TNI appears for the
ground-state magnetic moments of $^{11}$B and $^{11}$C.
Recently, the Gamow-Teller transition strength for the $^{11}$B$\rightarrow ^{11}$C
was measured from the $^{11}$B($p,n$)$^{11}$C reaction \cite{Tadd90}.
In Table \ref{tab_b11}, we summarize the experimental and our calculated Gamow-Teller 
transition results. Overall, much better agreement is obtained in the calculation with 
the TNI.

\subsection{$^{12}$C, $^{12}$B, $^{12}$N}\label{subs_c12}

\begin{figure}
\vspace*{2cm}
\includegraphics[width=8.0in]{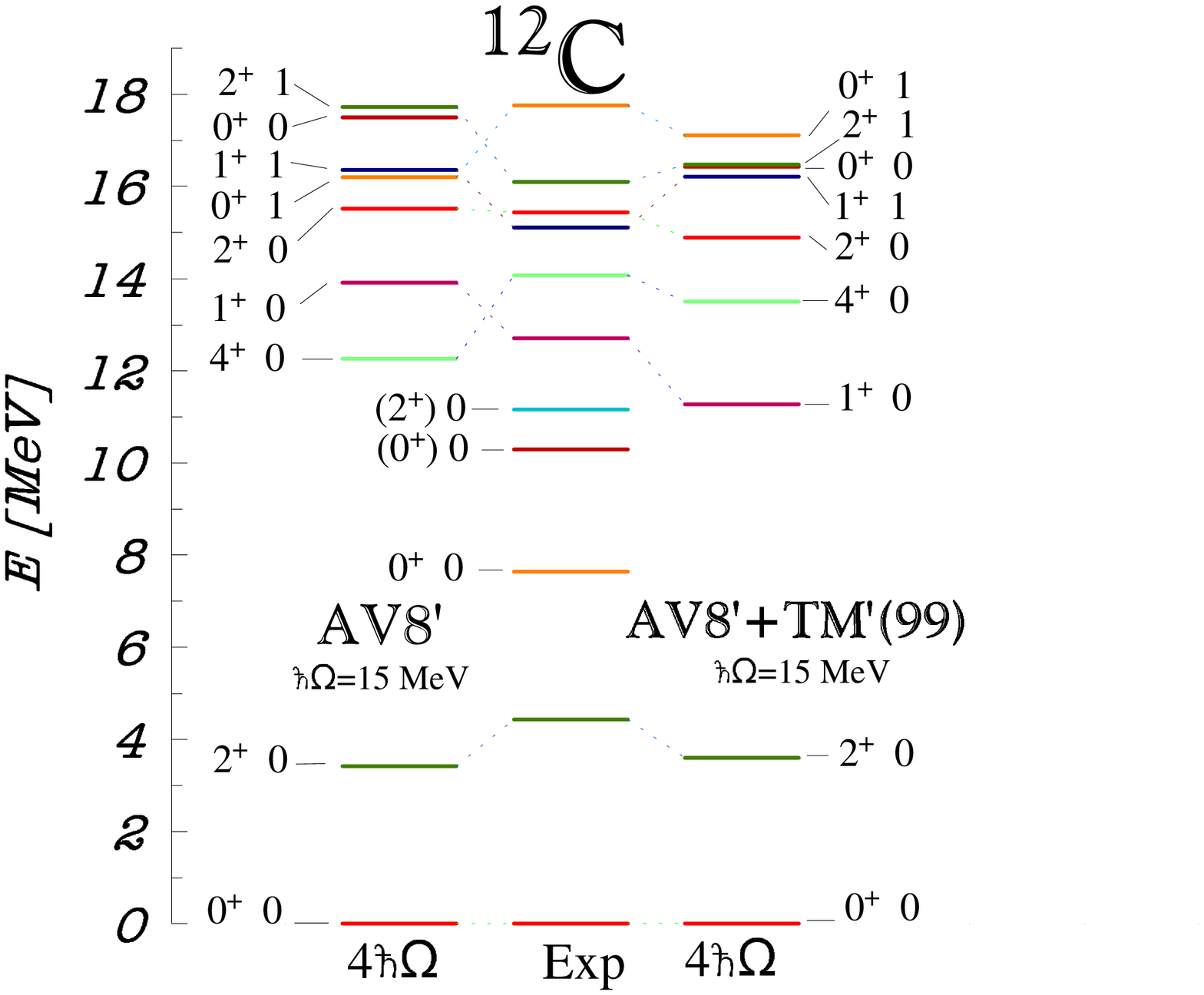}
\caption{\label{c12_exc_TM}Calculated positive-parity excitation spectra of
$^{12}$C obtained in $4\hbar\Omega$ basis space using three-body effective
interaction derived from AV8$^\prime$ NN potential and AV8$^\prime$ NN potential
plus TM$^\prime$(99) three-nucleon interaction, respectively,
are compared to experiment. The HO frequency of $\hbar\Omega=15$ MeV was used.
The experimental values are from Ref. \protect\cite{AS90}.
}
\end{figure}

Detailed $^{12}$C NCSM calculations using realistic two-nucleon interactions
were reported in Ref. \cite{C12_NCSM}. Here we extend 
those calculations by including the TNI. In Table \ref{tab_c12}, we summarize
some of our results. In Fig. \ref{c12_exc_TM}, we compare the excitation energies
obtained with and without the TNI.  
In addition to an increase of binding energy by 6 MeV, we observe a sensitivity 
of the low-lying spectra to the presence of the realistic three-body force and 
a trend toward level-ordering improvement in comparison to 
experiment. The sensitivity is the largest for states where the spin-orbit 
interaction strength is known to play a role. Note the correct ordering of the 
$1^+ 0 \leftrightarrow 4^+ 0$ states and ordering and spacing improvement 
of the lowest T=1 states. In fact, the lowest $T=1$ state, the isospin analog state 
of the $^{12}$N and $^{12}$B ground state comes out wrong, i.e., $0^+_1 1$ instead of
$1^+_1 1$, in the calculation with just the AV8$^\prime$. Once the TM$^\prime$(99)
is included we obtain the correct $1^+_1 1$ state as the lowest $T=1$ state
and, moreover, the correct ordering of the next two lowest $T=1$ states, $2^+_1 1$
and $0^+_1 1$.

We note that in Fig. \ref{c12_exc_TM} we also show low-lying $^{12}$C states that
are dominated by higher than $p$-shell configurations. We obtain such states 
at much higher excitation energies in the basis spaces employed in this investigation. 
 
The NCSM calculated binding energy of $^{12}$C decreases with the basis size 
enlargement \cite{C12_NCSM}. We extrapolate the binding energy for the
AV8$^\prime$ NN potential to be less than 80 MeV. Consequently, we expect
the binding energy for the AV8$^\prime$+TM$^\prime$(99) also decrease
if we were able to reach the $6\hbar\Omega$ or larger basis space.
Therefore, the good agreement of the AV8$^\prime$+TM$^\prime$(99) binding energy 
in the $4\hbar\Omega$ space with experiment shown in Table \ref{tab_c12} 
could be accidental. To have a better insight on this issue, we need to
perform more calculations, i.e., frequency and basis size dependence.  

\begin{figure}
\vspace*{2cm}
\includegraphics[width=8.0in]{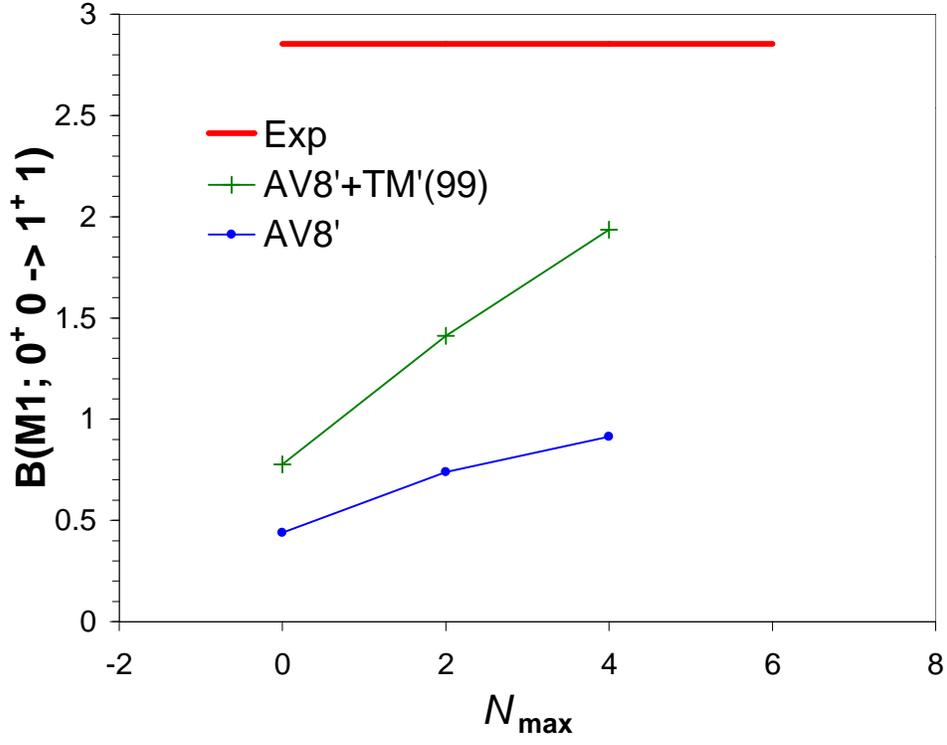}
\caption{\label{c12_bm1} Experimental and calculated B(M1) values, in $\mu_{\rm N}^2$,
for the $0^+ 0\rightarrow 1^+ 1$ transition in $^{12}$C. Results obtained
using the AV8$^\prime$ and AV8$^\prime$+TM$^\prime$(99) interactions
in basis spaces up to $4\hbar\Omega$ are compared. The HO frequency 
of $\hbar\Omega=15$ MeV was used. The experimental values are from 
Ref. \protect\cite{AS90}.
}
\end{figure}

It is well known that the isovector $M1$ transition from the $^{12}$C ground state 
to the 15.11 MeV $1^+_1 1$ state is
very sensitive to the strength of the spin orbit interaction.
Our B(M1; $0^+ 0\rightarrow 1^+ 1$) results are presented in Fig. \ref{c12_bm1}. 
The calculations with the 2-nucleon interaction show saturation and under-predict 
the experiment by almost a factor of three. By including the three-nucleon interaction, 
the B(M1) value increases dramatically. This improvement is almost entirely 
due to the improved strength of the spin-orbit splitting when the three-nucleon 
interaction is included. The basis size dependence is stronger in this case. 
Clearly, if we were able to extend our calculations to the
$6\hbar\Omega$ space, we would get much closer to the B(M1) experimental value.   
We also note a substantial increase and overall better agreement with experiment
for the B(E2; $2^+_1 1\rightarrow 0^+_1 0$) in the calculation with the TNI 
(see Table \ref{tab_c12}).
 
Our results for $^{12}$N and $^{12}$B presented in Table \ref{tab_c12}
further support the above discussed observations. In particular, the correct 
ground-state spin is obtained only in calculations that include the TNI.
There is a significant improvement in level spacing even for the higher
excited state, e.g., $2^+_2 1, 1^+_2 1$ and $3^+_1 1$.
It is interesting to note the complicated situation concerning the magnetic
moments of the $^{12}$N and $^{12}$B ground states. In the calculation,
there is a significant cancellation between orbital and spin contributions and, 
in particular, between proton and neutron spin contributions. Clearly, 
the calculations with the TNI improve agreement with experiment, although
a significant discrepancy remains.
The Gamow-Teller ground-state to ground-state transitions behave as the $M1$
isovector transition from the $^{12}$C ground state to the 15.11 MeV $1^+_1 1$ 
state. The improvement compared to experiment is quite dramatic when the TNI
is included.

\subsection{$^{13}$C}\label{subs_c13}

\begin{figure}
\vspace*{2cm}
\includegraphics[width=8.0in]{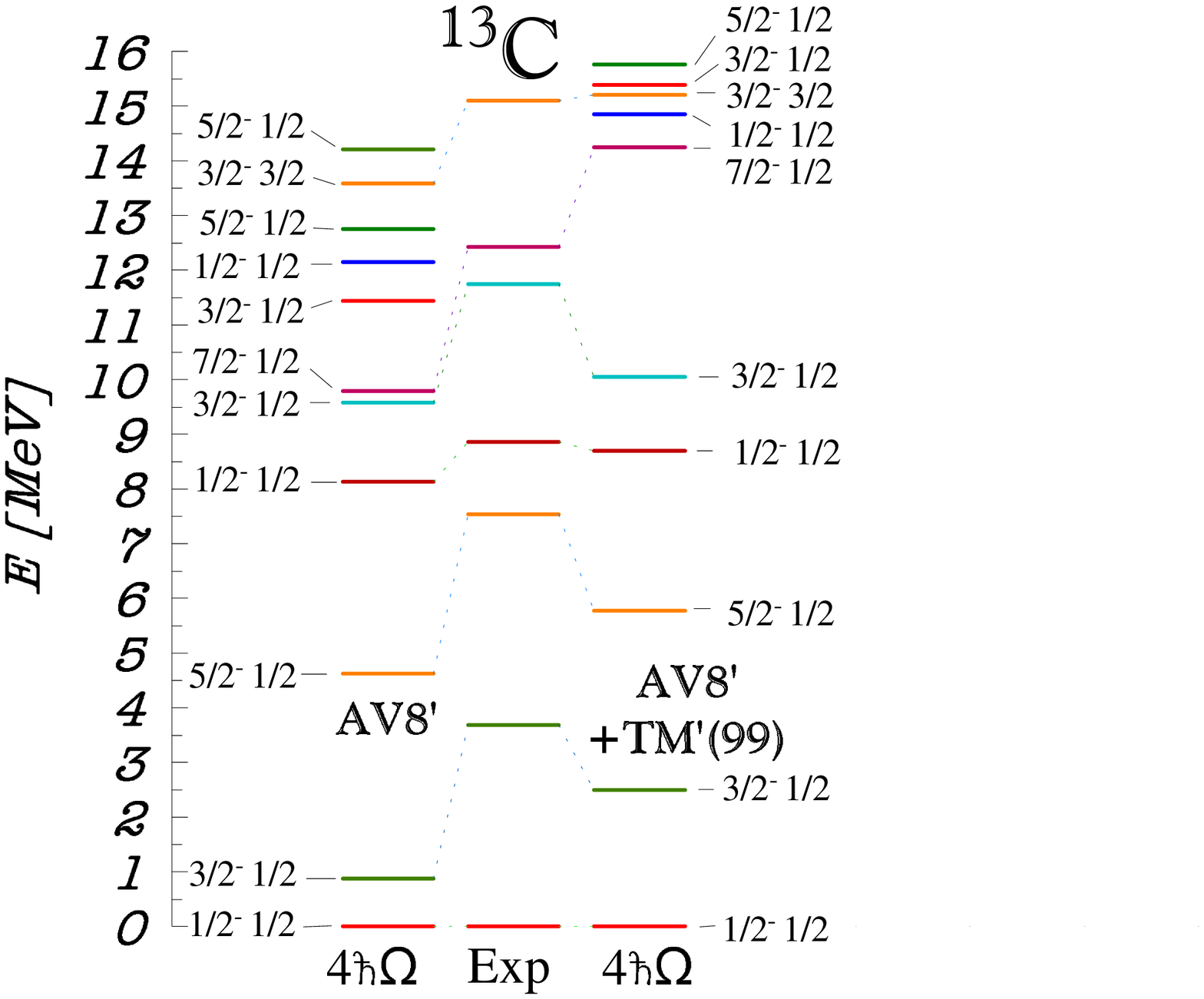}
\caption{\label{c13_exc_TM} Calculated negative-parity excitation spectra of
$^{13}$C obtained in $4\hbar\Omega$ basis space using three-body effective
interaction derived from AV8$^\prime$ NN potential and AV8$^\prime$ NN potential
plus TM$^\prime$(99) three-nucleon interaction, respectively,
are compared to experiment. The HO frequency of $\hbar\Omega=15$ MeV was used.
The level assignments are made according to Ref. \protect\cite{c13_millener}.
Only experimental states known to be dominantly of $p$-shell character are shown.
The experimental values are from Ref. \protect\cite{AS91}.
}
\end{figure}

NCSM calculations for $^{13}$C using a two-nucleon interaction were reported
in Ref. \cite{c13_scat}. In that work, the NCSM one-body transition densities
were used for the DWBA description of proton and $^3$He scattering on $^{13}$C.
Our present $^{13}$C results obtained with and without the TNI
are compared in Table \ref{tab_c13} and Fig. \ref{c13_exc_TM}.
We used the $4\hbar\Omega$ basis space using the same optimal frequency
as for $^{12}$C, i.e., $\hbar\Omega=15$ MeV.
The excitation spectrum of $^{13}$C is quite complex. As we discussed earlier 
our calculated states presented in this study are all $0\hbar\Omega$ dominated.
It is important to make a correct correspondence to the experimental 
levels as it is to be expected that the $sd$-shell dominated states would appear
at low energies in $^{13}$C. The character of low-lying $^{13}$C states was studied 
in Ref. \cite{c13_millener}. We use assignments found in Ref. \cite{c13_millener} 
in our Table \ref{tab_c13} and Fig. \ref{c13_exc_TM}.
Despite the complexity of the $^{13}$C excitation spectra the NCSM calculation
obtains the lowest six states in correct order in both calculations with and without
the TNI. When the TNI is included the lowest five states excitation energies
improve compared to experiment. Note in particular an increase of the excitation energy 
of the $\frac{3}{2}^-_1 \frac{1}{2}$ state. Also, the position of the lowest
$T=\frac{3}{2}$ state $\frac{3}{2}^-_1 \frac{3}{2}$ improves. 
From Table \ref{tab_c13} we can see that our AV8$^\prime$+TM$^\prime$(99)
$4\hbar\Omega$ calculation slightly overbinds $^{13}$C. Similar comment as
in Subsection \ref{subs_c12} on $^{12}$C applies. We expect the binding energy 
decrease with the basis enlargement.

\section{Effects of the Three-Nucleon Interaction}\label{TNI}

Overall, we find that three-nucleon interactions, and in particular the TM$^\prime$(99) force
employed in this work, do have an important effect on nuclear structure. Of course,
the primary expectation from previous studies of three- and four-nucleon systems 
with exact methods, is that the TNI will provide more binding energy in many-body systems. 
This requirement was certainly verified with previous NCSM and GFMC studies using 
realistic NN interactions. This is also borne out in the tables, and in the discussion 
that follows we focus primarily on the $A\le 10$ nuclei,
where the results with just the NN-interaction are reasonably converged, and are likely
to be within 1~MeV of the exact results~\cite{v3eff}.  
For $^6$Li, the NN-interaction accounts for nearly
89\% of the binding energy, while for $^7$Li, $^8$Be, and $^{10}$B it accounts for 84-86\% of
the total binding. The TM$^\prime$(99) force then contributes an additional binding of the 
order 8-10\% of the value obtained from the NN-interaction alone. What is somewhat 
remarkable is that the results with the TNI in $^6$Li come quite close to the 
experimental binding energy (underbinding by less than 1~MeV), while the TNI significantly 
underbinds the $A\ge 7$ nuclei by 3.5-4.2 MeV. It is interesting that the underbinding becomes
more dramatic with the addition just a single nucleon, i.e., from $^6$Li to $^7$Li. The
nature of the underbinding requires further research. We plan to address three
avenues of investigation. The first is the effects of the smaller three-particle model space
on the convergence of the three-body effective interaction, which may be more severe 
for smaller oscillator frequencies. Although, our expectation is that the full effect of 
the underbinding is not due to a convergence issue. Efforts to upgrade the computer 
programs used to evaluate $V_{{\rm 3eff},123}^{\rm NN+NNN}$ to use larger three-particle 
model spaces are underway. Second, we are also in the process of improving 
our shell-model codes to extend the calculations up to $N_{\rm max}=8$ for $A=6,7$ and
$N_{\rm max}=6$ for $8\le A\le 16$ nuclei. This extension is important in order to be able to 
fully assess the effect of the TNI on the structure of $6\le A \le 10$ nuclei. Finally, with 
these new capabilities, we hope to turn to the form of the TNI itself, which is not yet fully
determined.  
 
The other important feature of the TNI is the enhancement of spin-orbit effects. This is
particularly evident in $^{10}$B, where the inversion of the $1^+ 0$ and $3^+0$
states is a clear indication that the NN-interaction by itself is insufficient. 
A similar ground-state level inversions are observed for $^{11}$B, and $^{12}$N, $^{12}$B.
It is to be
noted that the correct ordering is obtained with a realistic TNI with no free 
parameters (other than a cutoff fixed by reproducing the $A=3$ binding energies). 
In addition, several calculated Gamow-Teller and $M1$ transitions are in 
better agreement with experiment when the TNI is included. These results are promising
in that the inclusion of the TNI can improve our overall picture of the structure of light
nuclei. 

\section{Conclusions}\label{sec_concl}

We extended the {\it ab initio} no-core shell model to include
a realistic three-body interaction in calculations for $p$-shell nuclei.
We presented results of first applications to $^{6,7}$Li, $^6$He, $^8$Be, $^{10,11,12}$B,
$^{12}$N and $^{10,12,13}$C using the Argonne V8$^\prime$ NN potential and 
the chiral-symmetry-based Tucson-Melbourne TM$^\prime$(99) three-nucleon interaction.
In general, 
in addition to increase of binding energy, we observed a substantial sensitivity 
of the low-lying spectra to the presence of the realistic three-nucleon interaction 
and a trend toward level-ordering and level-spacing improvement in comparison to 
experiment. The sensitivity is the largest for states where the spin-orbit 
interaction strength is known to play a role. In particular, with the TNI
we obtain the correct ground-state spin for $^{10,11}$B and $^{12}$N, $^{12}$B, 
contrary to calculations with NN potentials only. In addition, we observe an overall 
improvement of the Gamow-Teller transition description in calculations with the
three-nucleon interaction. The most striking case appears in $A=12$ nuclei
for the $0^+_1 0\rightarrow 1^+_1 1$ transitions. 

Unlike two-nucleon interaction, a detailed form and parametrization 
of the TNI is not well established yet, and it is highly likely that 
the sensitivity of the excitation spectra to the presence of the three-nucleon interaction
could be exploited in determining the structure of the three-nucleon interaction itself. 
This is a topic that we plan to pursue in the future.

Until now calculations with a realistic TNI in the $p$-shell has been performed
only using the GFMC method \cite{av8p,wiringa00,pieper01,GFMC_9_10} and by the coupled
cluster method for $^{16}$O \cite{CCM}. Our NCSM approach has some advantages compared
to those methods such as access to wider range of nuclei and also obtaining a more complete
set of excited states. An important issue for the NCSM is the convergence which
implies the need to use as large basis space as possible. With some algorithm and 
coding developments it should be feasible to perform $8\hbar\Omega$ calculations
with the TNI for the light $p$-shell nuclei and $6\hbar\Omega$ calculations for 
the rest of the $p$-shell in a near future. Another important topic is the
use of a greater variety of three-nucleon interactions and their different 
parametrizations including implementation of additional terms \cite{Bira,Friar} 
as well as the new effective field theory interactions developed in Ref. \cite{Bira1}
and applied in Ref. \cite{EFT_V3b}. A closely related issue is an improvement of efficiency
and accuracy of the three-nucleon system solution calculations needed 
to construct the effective interaction as described in step (i) of 
Section \ref{sec_theory}. A work in this direction is now under way \cite{Andreas}.

\section{Acknowledgments}

We thank S. A. Coon, A. Nogga and D. J. Millener for useful comments.
This work was performed under the auspices of
the U. S. Department of Energy by the University of California,
Lawrence Livermore National Laboratory under contract
No. W-7405-Eng-48.

\begin{table}
\begin{tabular}{c|c c c c c c} 
& $\mu a^\prime$ & $\mu^3 b$ & $\mu^3 d$ & $\Lambda$ [$\mu$] & $g^2$  & $\mu$ [MeV]\\
\hline 
TM$^\prime$(99) & -1.12 & -2.80 & -0.72 & 4.7 & 172.1 & 138.0\\
\hline
\end{tabular}
\caption{\label{TMparam} Constants of the TM$^\prime$(99) three-body force
used in the present investigation. 
}
\end{table}

\begin{table}
\begin{tabular}{c|c|c|c}
 $^{6}$Li & Exp & AV8$^\prime$+TM$^\prime$(99) & AV8$^\prime$    \\ 
 basis space & - &$6\hbar\Omega$ &$6\hbar\Omega$  \\
\hline
$|E_{\rm gs}|(1^+ 0)$      &  31.995 &  31.036 &  28.406 \\
$r_p$ [fm]                       &  2.32(3)  &   2.054 &   2.097 \\
$Q_{\rm gs}$ [$e$ fm$^2$]        & -0.082(2) &  -0.025 &  -0.102 \\ 
$\mu_{\rm gs}$ [$\mu_{\rm N}^2$] & +0.822    &  +0.840 &  +0.848 \\  
\hline
$E_{\rm x}(1^+_1 0)$ & 0.0    & 0.0    & 0.0   \\
$E_{\rm x}(3^+ 0)$   & 2.186  & 2.471  & 2.986 \\ 
$E_{\rm x}(0^+ 1)$   & 3.563  & 3.886  & 3.933 \\
$E_{\rm x}(2^+ 0)$   & 4.312 & 5.010  & 4.733 \\
$E_{\rm x}(2^+_1 1)$ & 5.366  & 6.482  & 6.563 \\
$E_{\rm x}(1^+_2 0)$ & 5.65   & 7.621  & 6.922 \\
$E_{\rm x}(2^+_2 1)$ &        & 10.693 & 10.095 \\
$E_{\rm x}(1^+ 1)$   &        & 11.525 & 10.881 \\
\hline
B(E2;$1^+_1 0 \rightarrow 3^+ 0$)   & 21.8(4.8)  &  7.093 &  8.412 \\
B(M1;$0^+ 1 \rightarrow 1^+_1 0$)   & 15.42(32)  & 15.374 & 15.234 \\
B(E2;$2^+ 0 \rightarrow 1^+_1 0$)   & 4.41(2.27) & 3.129  &  3.491 \\
B(M1;$2^+_1 1 \rightarrow 1^+_1 0$) & 0.150(27)  &  0.113 &  0.027 \\
\hline
\hline
 $^{6}$He & Exp & AV8$^\prime$+TM$^\prime$(99) & AV8$^\prime$    \\ 
 basis space & - &$6\hbar\Omega$ &$6\hbar\Omega$  \\
\hline
$|E_{\rm gs}|(0^+ 1)$ & 29.269  & 28.189 & 25.488 \\  
$r_p$ [fm]            & 1.72(4) &  1.707 &  1.743 \\
\hline
$E_{\rm x}(0^+_1 1)$ & 0.0  & 0.0    & 0.0   \\
$E_{\rm x}(2^+ 1)$   & 1.8  & 2.598  & 2.638 \\
$E_{\rm x}(2^+ 1)$   &      & 6.892  & 6.231 \\
$E_{\rm x}(1^+ 1)$   &      & 7.737  & 7.035 \\
$E_{\rm x}(0^+_2 1)$ &      & 10.855 & 9.467 \\
\hline
\hline
 $^{6}$He$\rightarrow ^6$Li & Exp & AV8$^\prime$+TM$^\prime$(99) & AV8$^\prime$ \\ 
 basis space & - &$6\hbar\Omega$ &$6\hbar\Omega$  \\
\hline
B(GT;$0^+_1 1\rightarrow 1^+_1 0$) & 4.728(15) & 5.213 & 5.314 \\
\end{tabular}
\caption{\label{tab_li6} Experimental and calculated energies, in MeV,
ground-state point-proton rms radii, quadrupole and magnetic moments,
as well as E2, in $e^2$ fm$^4$, and M1, in $\mu_N^2$, transitions
of $^{6}$Li and $^{6}$He as well as the B(GT)
values for the $^6$He ground state to $^6$Li ground state transition. 
Results obtained using three-body effective interactions derived from
the AV8$^\prime$ and AV8$^\prime$+TM$^\prime$(99) 
interactions are presented.
A HO frequency of $\hbar\Omega=14$ MeV was employed.
The experimental values are from Ref. \protect\cite{AS88,Till02,T88}.
}
\end{table}

\begin{table}
\begin{tabular}{c|c|c|c}
$^{7}$Li & Exp & AV8$^\prime$+TM$^\prime$(99) & AV8$^\prime$    \\ 
 basis space & - &$6\hbar\Omega$ &$6\hbar\Omega$  \\
\hline
$|E_{\rm gs}(\frac{3}{2}^- \frac{1}{2})|$ &  39.270 &  35.832 &  32.953 \\
$r_p$ [fm]                          & 2.27(2) &   2.071 &   2.104 \\
$Q_{\rm gs}$ [$e$ fm$^2$]           & -4.00(6)&  -2.563 &  -2.679 \\ 
$\mu_{\rm gs}$ [$\mu_{\rm N}^2$]    & +3.256  &  +3.025 &  +3.012 \\
\hline
$E_{\rm x}(\frac{3}{2}^-_1 \frac{1}{2})$ & 0.0    & 0.0    & 0.0   \\
$E_{\rm x}(\frac{1}{2}^-_1 \frac{1}{2})$ & 0.478  & 0.317  & 0.229 \\
$E_{\rm x}(\frac{7}{2}^-_1 \frac{1}{2})$ & 4.65   & 5.301  & 5.499 \\ 
$E_{\rm x}(\frac{5}{2}^-_1 \frac{1}{2})$ & 6.60   & 7.388  & 6.971 \\
$E_{\rm x}(\frac{5}{2}^-_2 \frac{1}{2})$ & 7.45   & 8.527  & 8.732 \\
$E_{\rm x}(\frac{3}{2}^-_2 \frac{1}{2})$ & 8.75   & 10.668 & 10.026 \\
$E_{\rm x}(\frac{1}{2}^-_2 \frac{1}{2})$ & 9.09   & 11.145 & 10.793 \\
$E_{\rm x}(\frac{7}{2}^-_2 \frac{1}{2})$ & 9.57   & 11.568 & 11.331 \\
$E_{\rm x}(\frac{3}{2}^-_1 \frac{3}{2})$ & 11.24  & 12.583 & 12.613 \\
\hline
\hline
$^{7}$Be & Exp & AV8$^\prime$+TM$^\prime$(99) & AV8$^\prime$    \\ 
 basis space & - &$6\hbar\Omega$ &$6\hbar\Omega$  \\
\hline
$|E_{\rm gs}(\frac{3}{2}^- \frac{1}{2})|$ &  37.600 & 34.176 & 31.318 \\
$r_p$ [fm]                          &  2.36(2)   & 2.225  &  2.262 \\
$Q_{\rm gs}$ [$e$ fm$^2$]           &            & -4.305 & -4.614 \\
$\mu_{\rm gs}$ [$\mu_{\rm N}^2$]    & -1.398(15) & -1.153 & -1.138 \\
\hline
$E_{\rm x}(\frac{3}{2}^-_1 \frac{1}{2})$ & 0.0      & 0.0    & 0.0   \\
$E_{\rm x}(\frac{1}{2}^-_1 \frac{1}{2})$ & 0.429    & 0.309  & 0.225 \\
$E_{\rm x}(\frac{7}{2}^-_1 \frac{1}{2})$ & 4.57(5)  & 5.232  & 5.429 \\
$E_{\rm x}(\frac{5}{2}^-_1 \frac{1}{2})$ & 6.73(10) & 7.329  & 6.912 \\
$E_{\rm x}(\frac{5}{2}^-_2 \frac{1}{2})$ & 7.21(6)  & 8.333  & 8.514 \\
\hline
\hline
$^{7}$Be$\rightarrow ^7$Li & Exp & AV8$^\prime$+TM$^\prime$(99) & AV8$^\prime$ \\ 
 basis space & - &$6\hbar\Omega$ &$6\hbar\Omega$  \\
\hline
B(GT;$\frac{3}{2}^-_1 \frac{1}{2}\rightarrow \frac{3}{2}^-_1 \frac{1}{2}$) &  1.300 & 1.401 & 1.445 \\
B(GT;$\frac{3}{2}^-_1 \frac{1}{2}\rightarrow \frac{1}{2}^-_1 \frac{1}{2}$) &  1.122 & 1.214 & 1.231 \\
\end{tabular}
\caption{\label{tab_li7} Experimental and calculated energies, in MeV,
ground-state point-proton rms radii, quadrupole and magnetic 
moments,
of $^{7}$Li and $^{7}$Be as well as the B(GT) values for the $^7$Be 
to $^7$Li transitions. 
Results obtained using three-body effective interactions derived from
the AV8$^\prime$ and AV8$^\prime$+TM$^\prime$(99) 
interactions are presented.
A HO frequency of $\hbar\Omega=14$ MeV was employed.
The experimental values are from Ref. \protect\cite{Till02,T88,Chou93}.
}
\end{table}

\begin{table}
\begin{tabular}{c|c|c|c}
$^{8}$Be & Exp & AV8$^\prime$+TM$^\prime$(99) & AV8$^\prime$    \\ 
 basis space & - &$4\hbar\Omega$ &$4\hbar\Omega$  \\
\hline
$|E_{\rm gs}|$   & 56.50 & 52.328 & 48.454 \\
\hline
$E_{\rm x}(0^+_1 0)$ & 0.0   & 0.0    & 0.0    \\
$E_{\rm x}(2^+_1 0)$ & 3.04  & 3.724  & 3.652  \\
$E_{\rm x}(4^+_1 0)$ & 11.40 & 12.713 & 12.402 \\
$E_{\rm x}(2^+_1 1)^a$ & 16.63$^a$ & 16.831$^a$ & 16.630 \\
$E_{\rm x}(2^+_2 0)^a$ & 16.92$^a$ & 17.009$^a$ & 16.998 \\
$E_{\rm x}(1^+_1 1)$ & 17.64 & 18.049 & 17.649 \\
$E_{\rm x}(1^+_1 0)$ & 18.15 & 17.895 & 17.463 \\
$E_{\rm x}(3^+_1 1)$ & 19.01 & 19.797 & 19.608 \\
$E_{\rm x}(3^+_1 0)$ & 19.24 & 20.163 & 19.028 \\
$E_{\rm x}(0^+_1 1)$ &       & 20.910 & 19.525 \\
$E_{\rm x}(4^+_2 0)$ & 19.86 & 20.942 & 21.144 \\
$E_{\rm x}(0^+_2 0)$ &       & 27.705 & 26.478 \\
$E_{\rm x}(0^+_1 2)$ & 27.49 & 28.931 & 28.335 \\
\hline
\end{tabular}
\vspace{2mm}
\newline
\noindent\small{$^a$T=0 and T=1 components strongly mixed}
\caption{\label{tab_be8} Experimental and calculated energies, in MeV,
of $^{8}$Be. Results obtained 
using three-body effective interactions derived from
the AV8$^\prime$ and AV8$^\prime$+TM$^\prime$(99) 
interactions are presented.
A HO frequency of $\hbar\Omega=14$ MeV was employed.
The experimental values are from Ref. \protect\cite{AS88}.
}
\end{table}

\begin{table}
\begin{tabular}{c|c|c|c}
$^{10}$B & Exp & AV8$^\prime$+TM$^\prime$(99) & AV8$^\prime$    \\ 
 basis space & - &$4\hbar\Omega$ &$4\hbar\Omega$  \\
\hline
$|E (3^+ 0)|$       &  64.751     & 60.567 & 54.833 \\
$r_p(3^+ 0)$ [fm]       &  2.30(12)   &  2.168 & 2.196  \\ 
$Q_{3^+ 0}$ [$e$ fm$^2$] &  +8.472(56) & +5.682 & +5.937 \\
$\mu_{3^+ 0}$ [$\mu_{\rm N}^2$] & +1.8006   & +1.847 & +1.857 \\
$|E (1^+ 0)|$       &  64.033     & 60.227 & 55.979 \\
$\mu_{1^+ 0}$ [$\mu_{\rm N}^2$] & +0.63(12) & +0.802 & +0.843 \\
\hline
$E_{\rm x}(3^+_1 0)$  & 0.0   & 0.0    &  0.0   \\
$E_{\rm x}(1^+_1 0)$  & 0.718 & 0.340  & -1.146 \\
$E_{\rm x}(0^+_1 1)$  & 1.740 & 1.259  &  1.039 \\
$E_{\rm x}(1^+_2 0)$  & 2.154 & 1.216  &  1.664 \\
$E_{\rm x}(2^+_1 0)$  & 3.587 & 2.775  &  1.579 \\
$E_{\rm x}(3^+_2 0)$  & 4.774 & 5.971  &  4.363 \\
$E_{\rm x}(2^+_1 1)$  & 5.164 & 5.182  &  4.553 \\
$E_{\rm x}(2^+_2 0)$  & 5.92  & 3.987  &  3.470 \\
$E_{\rm x}(4^+_1 0)$  & 6.025 & 5.229  &  4.732 \\
$E_{\rm x}(2^+_2 1)$  & 7.478 & 7.491  &  5.741 \\
\hline
B(E2;$1^+_1 0 \rightarrow 3^+_1 0$) & 4.13(6) & 1.959 & 3.568 \\
B(E2;$1^+_2 0 \rightarrow 3^+_1 0$) & 1.71(26)& 1.010 & 0.047 \\
B(E2;$1^+_2 0 \rightarrow 1^+_1 0$) & 0.83(40)& 3.384 & 2.311 \\
B(E2;$3^+_2 0 \rightarrow 1^+_1 0$) & 20.5(26)& 3.543 & 3.289 \\
\hline
\hline
$^{10}$Be & Exp & AV8$^\prime$+TM$^\prime$(99) & AV8$^\prime$    \\ 
 basis space & - &$4\hbar\Omega$ &$4\hbar\Omega$  \\
$|E_{\rm gs}|$   & 64.977   &  61.387 &  55.840 \\ 
$r_p$ [fm]       &  2.24(8) &   2.087 &   2.113 \\
\hline
$E_{\rm x}(0^+_1 1)$  & 0.0   & 0.0    &  0.0   \\
$E_{\rm x}(2^+_1 1)$  & 3.368 &  3.877 &  3.463 \\
$E_{\rm x}(2^+_2 1)$  & 5.958 &  6.241 &  4.706 \\
$E_{\rm x}(1^+_1 1)$  &       &  8.532 &  7.582 \\
$E_{\rm x}(3^+_1 1)$  &       &  9.856 &  8.190 \\
$E_{\rm x}(2^+_3 1)$  & 9.4   & 10.036 &  9.040 \\
\hline
\hline
$^{10}$B$\rightarrow ^{10}$Be & Exp & AV8$^\prime$+TM$^\prime$(99) & AV8$^\prime$\\ 
 basis space & - &$4\hbar\Omega$ &$4\hbar\Omega$  \\
\hline
B(GT;$3^+_1 0\rightarrow 2^+_1 1$) & 0.08(3)  & 0.066 &  0.062 \\ 
B(GT;$3^+_1 0\rightarrow 2^+_2 1$) & 0.95(13) & 1.291 &  1.554 \\
\hline
\hline
$^{10}$C & AV8$^\prime$+TM$^\prime$(99) & Exp & AV8$^\prime$    \\ 
 basis space & - &$4\hbar\Omega$ &$4\hbar\Omega$  \\
$|E_{\rm gs}|$   & 60.321   & 56.626 &  51.141 \\
$r_p$ [fm]       &  2.31(3) &  2.246 &   2.279 \\
\hline
$E_{\rm x}(0^+_1 1)$  & 0.0   & 0.0    &  0.0   \\
$E_{\rm x}(2^+_1 1)$  & 3.354 & 3.913  &  3.508 \\
$E_{\rm x}(2^+_2 1)$  &       & 6.133  &  4.612 \\
$E_{\rm x}(1^+_1 1)$  &       & 8.437  &  7.453 \\
$E_{\rm x}(3^+_1 1)$  &       & 9.773  &  8.145 \\ 
$E_{\rm x}(2^+_3 1)$  &       & 10.049 &  9.009 \\
\hline
\hline
less $^{10}$C$\rightarrow ^{10}$B & Exp & AV8$^\prime$+TM$^\prime$(99) & AV8$^\prime$\\ 
 basis space & - &$4\hbar\Omega$ &$4\hbar\Omega$  \\
\hline
B(GT;$0^+_1 1\rightarrow 1^+_1 0$) & 3.44 & 4.331 & 4.748 \\ 
\end{tabular}
\caption{\label{tab_b10} Experimental and calculated energies, in MeV,
ground-state point-proton rms radii, 
the quadrupole moments,
as well as E2, in $e^2$ fm$^4$, transitions
of $^{10}$B, $^{10}$Be and $^{10}$C as well as selected Gamow-Teller transitions. 
Results obtained using three-body effective interactions derived from
the AV8$^\prime$ and AV8$^\prime$+TM$^\prime$(99) 
interactions are presented.
A HO frequency of $\hbar\Omega=15$ MeV was employed.
The experimental values are from Ref. \protect\cite{AS88,T88,Daito98,Chou93}.
}
\end{table}

\begin{table}
\begin{tabular}{c|c|c|c}
$^{11}$B & Exp & AV8$^\prime$+TM$^\prime$(99) & AV8$^\prime$    \\ 
 basis space & - &$4\hbar\Omega$ &$4\hbar\Omega$  \\
\hline
$|E(\frac{3}{2}^-_1 \frac{1}{2})|$              & 76.205    & 73.338  & 67.214 \\
$r_p(\frac{3}{2}^-_1 \frac{1}{2})$ [fm]             &  2.24(12)  &  2.148  & 2.175  \\
$Q(\frac{3}{2}^-_1 \frac{1}{2})$ [$e$ fm$^2$]        & +4.065(26)& +2.920  & +2.674 \\
$\mu(\frac{3}{2}^-_1 \frac{1}{2})$ [$\mu_{\rm N}^2$] & +2.689    & +2.176  & +2.708 \\
$|E(\frac{1}{2}^-_1 \frac{1}{2})|$              & 74.080    & 72.816  & 67.413 \\
$\mu(\frac{1}{2}^-_1 \frac{1}{2})$ [$\mu_{\rm N}^2$] &           & -0.435  & -0.505 \\
\hline
$E_{\rm x}(\frac{3}{2}^-_1 \frac{1}{2})$ & 0.0    & 0.0    & 0.0   \\
$E_{\rm x}(\frac{1}{2}^-_1 \frac{1}{2})$ & 2.125  & 0.522  &-0.198 \\ 
$E_{\rm x}(\frac{5}{2}^-_1 \frac{1}{2})$ & 4.445  & 3.565  & 2.642 \\
$E_{\rm x}(\frac{3}{2}^-_2 \frac{1}{2})$ & 5.020  & 3.840  & 2.492 \\
$E_{\rm x}(\frac{7}{2}^-_1 \frac{1}{2})$ & 6.743  & 5.481  & 4.946 \\
$E_{\rm x}(\frac{5}{2}^-_2 \frac{1}{2})$ & 8.92   & 9.141  & 8.130 \\
$E_{\rm x}(\frac{1}{2}^-_2 \frac{1}{2})$ &        & 9.819  & 10.372 \\
$E_{\rm x}(\frac{3}{2}^-_3 \frac{1}{2})$ &        & 10.643 & 10.280 \\
$E_{\rm x}(\frac{5}{2}^-_3 \frac{1}{2})$ & 11.64  & 11.989 &  9.572 \\
$E_{\rm x}(\frac{3}{2}^-_4 \frac{1}{2})$ &        & 12.839 & 12.264 \\
$E_{\rm x}(\frac{1}{2}^-_1 \frac{3}{2})$ & 12.916 & 13.479 & 12.443 \\
$E_{\rm x}(\frac{9}{2}^-_1 \frac{1}{2})$ &        & 13.402 & 11.139 \\
$E_{\rm x}(\frac{7}{2}^-_2 \frac{1}{2})$ &        & 14.026 & 12.143 \\
$E_{\rm x}(\frac{3}{2}^-_1 \frac{3}{2})$ & 15.29  & 16.588 & 14.735 \\
$E_{\rm x}(\frac{5}{2}^-_1 \frac{3}{2})$ & 16.50  & 17.056 & 15.968 \\ 
\hline
B(E2;$\frac{3}{2}^-_1 \frac{1}{2}\rightarrow \frac{1}{2}^-_1 \frac{1}{2} $) & 2.6(4) & 1.141 & 0.522 \\   
\hline
\hline
$^{11}$C & Exp & AV8$^\prime$+TM$^\prime$(99) & AV8$^\prime$    \\ 
 basis space & - &$4\hbar\Omega$ &$4\hbar\Omega$  \\
\hline
$|E(\frac{3}{2}^-_1 \frac{1}{2})|$                   &   73.440  & 70.618 & 64.515 \\
$Q(\frac{3}{2}^-_1 \frac{1}{2})$ [$e$ fm$^2$]        & +3.327(24)& +2.363 & +1.627 \\   
$\mu(\frac{3}{2}^-_1 \frac{1}{2})$ [$\mu_{\rm N}^2$] & -0.964(1) & -0.460 & -0.923 \\
$|E(\frac{1}{2}^-_1 \frac{1}{2})|$                   &   71.440  & 70.093 & 64.712 \\
$\mu(\frac{1}{2}^-_1 \frac{1}{2})$ [$\mu_{\rm N}^2$] &           & +0.809 & +0.882 \\
\hline
$E_{\rm x}(\frac{3}{2}^-_1 \frac{1}{2})$ & 0.0    & 0.0    &  0.0   \\
$E_{\rm x}(\frac{1}{2}^-_1 \frac{1}{2})$ & 2.000  & 0.525  & -0.197 \\
$E_{\rm x}(\frac{5}{2}^-_1 \frac{1}{2})$ & 4.319  & 3.584  &  2.656 \\
$E_{\rm x}(\frac{3}{2}^-_2 \frac{1}{2})$ & 4.804  & 3.852  &  2.498 \\
$E_{\rm x}(\frac{7}{2}^-_1 \frac{1}{2})$ & 6.478  & 5.363  &  4.848 \\
$E_{\rm x}(\frac{5}{2}^-_2 \frac{1}{2})$ & 8.420  & 8.943  &  7.978 \\
\hline
\hline
$^{11}$B$\rightarrow ^{11}$C & Exp & AV8$^\prime$+TM$^\prime$(99) & AV8$^\prime$    \\ 
 basis space & - &$4\hbar\Omega$ &$4\hbar\Omega$  \\
\hline
B(GT;$\frac{3}{2}^-_1 \frac{1}{2}\rightarrow \frac{3}{2}^-_1 \frac{1}{2}$) &  0.345 & 0.315 & 0.765 \\
B(GT;$\frac{3}{2}^-_1 \frac{1}{2}\rightarrow \frac{1}{2}^-_1 \frac{1}{2}$) &  0.399 & 0.591 & 0.909 \\
B(GT;$\frac{3}{2}^-_1 \frac{1}{2}\rightarrow \frac{5}{2}^-_1 \frac{1}{2}$) &  0.961$^a$ & 0.517 & 0.353 \\
B(GT;$\frac{3}{2}^-_1 \frac{1}{2}\rightarrow \frac{3}{2}^-_2 \frac{1}{2}$) &  0.961$^a$ & 0.741 & 0.531 \\
B(GT;$\frac{3}{2}^-_1 \frac{1}{2}\rightarrow \frac{5}{2}^-_2 \frac{1}{2}$) &  0.444$^b$ & 0.625 & 0.197 \\
\hline
\end{tabular}
\vspace{2mm}
\newline
\noindent\small{$^a$Unresolved doublet, $E_{\rm x}=4.32+4.80$ MeV, approximately equal strength.}
\newline
\noindent\small{$^b$Unresolved doublet, $E_{\rm x}=8.10+8.42$ MeV, most strength in the 8.42-MeV transition.}
\caption{\label{tab_b11} Experimental and calculated energies, in MeV, 
ground-state point-proton rms radii, quadrupole and magnetic moments,
E2, in $e^2$ fm$^4$,
transitions of $^{11}$B and $^{11}$C as well as the B(GT) values for the $^{11}$B 
to $^{11}$C transitions.  
Results obtained using three-body effective interactions derived from
the AV8$^\prime$ and AV8$^\prime$+TM$^\prime$(99) 
interactions are presented.
A HO frequency of $\hbar\Omega=15$ MeV was employed.
Only experimental states known to be dominantly of $p$-shell character 
are shown.
The experimental values are from Ref. \protect\cite{T88,AS90,Tadd90,Arya85}.
}
\end{table}

\begin{table}
\begin{tabular}{c|c|c|c}
$^{12}$C & Exp & AV8$^\prime$+TM$^\prime$(99) & AV8$^\prime$    \\ 
 basis space & - &$4\hbar\Omega$ &$4\hbar\Omega$  \\
\hline
$|E_{\rm gs}|$   & 92.162 & 91.963 & 85.945   \\
$r_p$ [fm]             & 2.35(2)& 2.191 & 2.209     \\
$Q_{2^+}$ [$e$ fm$^2$] & +6(3)  & 4.288 & 4.613     \\
\hline
$E_{\rm x}(0^+ 0)$ & 0.0    & 0.0    & 0.0      \\
$E_{\rm x}(2^+ 0)$ & 4.439  & 3.603  & 3.427   \\
$E_{\rm x}(1^+ 0)$ & 12.710 & 11.280 & 13.926 \\
$E_{\rm x}(4^+ 0)$ & 14.083 & 13.517 & 12.272 \\
$E_{\rm x}(1^+ 1)$ & 15.110 & 16.221 & 16.364 \\
$E_{\rm x}(2^+ 1)$ & 16.106 & 16.467 & 17.712 \\
$E_{\rm x}(0^+ 1)$ & 17.760 & 17.116 & 16.213 \\
\hline
B(E2;$2^+0 \rightarrow 0^+0$) & 7.59(42)  & 4.146  & 4.765  \\
B(M1;$1^+1 \rightarrow 0^+0$) & 0.951(20) & 0.645  & 0.305  \\
B(E2;$2^+1 \rightarrow 0^+0$) & 0.65(13)  & 0.430  & 0.247\\
\end{tabular}
\caption{\label{tab_c12} Experimental and calculated energies, in MeV,
the $2^+_1$-state quadrupole moments,
as well as E2, in $e^2$ fm$^4$, and M1, in $\mu_N^2$, transitions
of $^{12}$C. Results obtained 
using three-body effective interactions derived from
the AV8$^\prime$ and AV8$^\prime$+TM$^\prime$(99) 
interactions are presented.
A HO frequency of $\hbar\Omega=15$ MeV was employed.
The experimental values are from Ref. \protect\cite{AS90,T88}.
}
\end{table}

\begin{table}
\begin{tabular}{c|c|c|c}
$^{12}$B & Exp & AV8$^\prime$+TM$^\prime$(99) & AV8$^\prime$ \\ 
 basis space & - &$4\hbar\Omega$ &$4\hbar\Omega$  \\
\hline
$|E (1^+_1 1)|$            &  79.578   & 78.379  & 72.209 \\
$Q_{1^+_1 1}$ [$e$ fm$^2$] &  1.34(14) & +0.692  & +0.772 \\
$\mu_{1^+_1 1}$ [$\mu_{\rm N}^2$] & +1.003 & +0.2917 & -0.130 \\
\hline
$E_{\rm x}(1^+_1 1)$ & 0.0      & 0.0    &  0.0   \\
$E_{\rm x}(2^+_1 1)$ & 0.953    & 0.216  &  1.335 \\ 
$E_{\rm x}(0^+_1 1)$ & 2.723    & 0.854  & -0.158 \\
$E_{\rm x}(2^+_2 1)$ & 3.759(6) & 3.124  &  1.533 \\
$E_{\rm x}(1^+_2 1)$ & 5.00(20) & 4.381  &  2.727 \\
$E_{\rm x}(3^+_1 1)$ & 5.612(8) & 5.205  &  3.630 \\
\hline
\hline
 $^{12}$C$\rightarrow ^{12}$B & Exp & AV8$^\prime$+TM$^\prime$(99) & AV8$^\prime$ \\ 
 basis space & - &$4\hbar\Omega$ &$4\hbar\Omega$  \\
\hline
B(GT;$0^+_1 0\rightarrow 1^+_1 1$) & 0.990(2) & 0.666 & 0.255 \\
\hline
\hline
 $^{12}$N & Exp & AV8$^\prime$+TM$^\prime$(99) & AV8$^\prime$    \\ 
 basis space & - &$4\hbar\Omega$ &$4\hbar\Omega$  \\
\hline
$|E (1^+_1 1)|$             & 74.041   &  72.601 & 66.466 \\
$Q_{1^+_1 1}$ [$e$ fm$^2$]  & +1.03(7) &  +0.390 & +0.386 \\
$\mu_{1^+_1 1}$ [$\mu_{\rm N}^2$] & +0.457 & +1.124 & +1.513 \\
\hline
$E_{\rm x}(1^+_1 1)$ & 0.0       & 0.0    &  0.0   \\
$E_{\rm x}(2^+_1 1)$ & 0.960(12) & 0.236  &  1.338 \\ 
$E_{\rm x}(0^+_1 1)$ & 2.439(9)  & 0.825  & -0.167 \\
$E_{\rm x}(2^+_2 1)$ &           & 3.104  &  1.534 \\
$E_{\rm x}(1^+_2 1)$ &           & 4.392  &  2.733 \\
$E_{\rm x}(3^+_1 1)$ &           & 5.077  &  3.547 \\
\hline
\hline
$^{12}$C$\rightarrow ^{12}$N & Exp & AV8$^\prime$+TM$^\prime$(99) & AV8$^\prime$ \\ 
 basis space & - &$4\hbar\Omega$ &$4\hbar\Omega$  \\
\hline
B(GT;$0^+_1 0\rightarrow 1^+_1 1$) & 0.877(3) & 0.662 & 0.254 \\
\end{tabular}
\caption{\label{tab_b12_n12} Experimental and calculated energies, in MeV,
quadrupole and magnetic moments
of $^{12}$B and $^{12}$N as well as the B(GT)
values for the $^{12}$C ground state to $^{12}$B and $^{12}$N ground state transition.  
Results obtained using three-body effective interactions derived from
the AV8$^\prime$ and AV8$^\prime$+TM$^\prime$(99) 
interactions are presented.
A HO frequency of $\hbar\Omega=15$ MeV was employed.
The experimental values are from Ref. \protect\cite{AS90,Chou93}.
}
\end{table}

\begin{table}
\begin{tabular}{c|c|c|c}
$^{13}$C & Exp & AV8$^\prime$+TM$^\prime$(99) & AV8$^\prime$    \\ 
 basis space & - &$4\hbar\Omega$ &$4\hbar\Omega$  \\
\hline
$|E_{\rm gs}|$             & 97.108 & 99.220 & 91.889 \\
$r_p$ [fm]                       & 2.29(3) & 2.170  & 2.190  \\
$\mu_{\rm gs}$ [$\mu_{\rm N}^2$] & +0.702 & +0.630 & +0.911 \\ 
\hline
$E_{\rm x}(\frac{1}{2}^-_1 \frac{1}{2})$ & 0.0    & 0.0    & 0.0   \\
$E_{\rm x}(\frac{3}{2}^-_1 \frac{1}{2})$ & 3.685  & 2.501  & 0.886 \\
$E_{\rm x}(\frac{5}{2}^-_1 \frac{1}{2})$ & 7.547  & 5.781  & 4.628 \\
$E_{\rm x}(\frac{1}{2}^-_2 \frac{1}{2})$ & 8.860  & 8.702  & 8.135 \\
$E_{\rm x}(\frac{3}{2}^-_2 \frac{1}{2})$ & 11.748 & 10.054 & 9.588 \\
$E_{\rm x}(\frac{7}{2}^-_1 \frac{1}{2})$ & 12.438 & 14.244 & 9.795 \\
$E_{\rm x}(\frac{1}{2}^-_3 \frac{1}{2})$ &        & 14.856 & 12.151\\
$E_{\rm x}(\frac{3}{2}^-_1 \frac{3}{2})$ & 15.108 & 15.210 & 13.592\\
$E_{\rm x}(\frac{3}{2}^-_3 \frac{1}{2})$ &        & 15.396 & 11.441\\
$E_{\rm x}(\frac{5}{2}^-_2 \frac{1}{2})$ &        & 15.764 & 12.756\\
$E_{\rm x}(\frac{5}{2}^-_3 \frac{1}{2})$ &        & 18.387 & 14.207\\
\hline
B(E2;$ \frac{3}{2}^-_1 \frac{1}{2} \rightarrow \frac{1}{2}^-_1 \frac{1}{2}$) & 6.4(15) & 2.815 & 4.224 \\
B(M1;$ \frac{3}{2}^-_1 \frac{1}{2} \rightarrow \frac{1}{2}^-_1 \frac{1}{2}$) & 0.70(7) & 0.806 & 1.213 \\
\end{tabular}
\caption{\label{tab_c13} Experimental and calculated energies, in MeV,
ground-state point-proton rms radii, magnetic 
moments,
as well as E2, in $e^2$ fm$^4$, and M1, in $\mu_N^2$, transitions
of $^{13}$C. Results obtained 
using three-body effective interactions derived from
the AV8$^\prime$ and AV8$^\prime$+TM$^\prime$(99) 
interactions are presented.
A HO frequency of $\hbar\Omega=15$ MeV was employed.
Only experimental states known to be dominantly of $p$-shell character 
\protect\cite{c13_millener} are shown.
The experimental values are from Ref. \protect\cite{AS91,T88}.
}
\end{table}


\begin{thebibliography}{10}
\expandafter\ifx\csname bibnamefont\endcsname\relax
  \def\bibnamefont#1{#1}\fi
\expandafter\ifx\csname bibfnamefont\endcsname\relax
  \def\bibfnamefont#1{#1}\fi
\expandafter\ifx\csname url\endcsname\relax
  \def\url#1{\texttt{#1}}\fi
\expandafter\ifx\csname urlprefix\endcsname\relax\def\urlprefix{URL }\fi
\providecommand{\bibinfo}[2]{#2}
\providecommand{\eprint}[2][]{\url{#2}}


\bibitem{C12_NCSM}
\bibinfo{author}{\bibnamefont{{P. Navr\'atil, J. P. Vary and B. R. Barrett}}},
 \bibinfo{journal}{Phys. Rev. Lett.} \textbf{\bibinfo{volume}{84}},
  \bibinfo{pages}{5728} (\bibinfo{year}{2000});
  \bibinfo{journal}{Phys. Rev. C} \textbf{\bibinfo{volume}{62}},
  \bibinfo{pages}{054311} (\bibinfo{year}{2000}).

\bibitem{fourb_NCSM}
\bibinfo{author}{\bibnamefont{{P. Navr\'atil and B. R. Barrett}}}, 
\bibinfo{journal}{Phys. Rev. C} \textbf{\bibinfo{volume}{59}},
  \bibinfo{pages}{1906} (\bibinfo{year}{1999}).

\bibitem{Jacobi_NCSM}
\bibinfo{author}{\bibnamefont{{P. Navr\'atil, G. P. Kamuntavi\v{c}ius and B. R.
  Barrett}}}, \bibinfo{journal}{Phys. Rev. C} \textbf{\bibinfo{volume}{61}},
  \bibinfo{pages}{044001} (\bibinfo{year}{2000}).

\bibitem{v3eff}
\bibinfo{author}{\bibfnamefont{P.}~\bibnamefont{Navr\'atil}} \bibnamefont{and}
  \bibinfo{author}{\bibfnamefont{W.~E.} \bibnamefont{Ormand}},
  \bibinfo{journal}{Phys. Rev. Lett.} \textbf{\bibinfo{volume}{88}},
  \bibinfo{pages}{152502} (\bibinfo{year}{2002}).

\bibitem{NCSM_TM}
\bibinfo{author}{\bibnamefont{{D. C. J. Marsden, P. Navr\'atil, S. A. Coon 
                 and B. R. Barrett}}},
  \bibinfo{journal}{Phys. Rev. C} \textbf{\bibinfo{volume}{66}},
  \bibinfo{pages}{044007} (\bibinfo{year}{2002}).

\bibitem{av8p}
\bibinfo{author}{\bibfnamefont{B. S. Pudliner, V. R. Pandharipande, J. Carlson, 
               S. C. Pieper and R. B. Wiringa}},
  \bibinfo{journal}{Phys. Rev. C} \textbf{\bibinfo{volume}{56}},
  \bibinfo{pages}{1720} (\bibinfo{year}{1997}).

\bibitem{wiringa00}
\bibinfo{author}{\bibnamefont{{R. B. Wiringa, S. C. Pieper, J. Carlson, V. R.
  Pandharipande}}}, \bibinfo{journal}{Phys. Rev. C}
  \textbf{\bibinfo{volume}{62}}, \bibinfo{pages}{014001}
  (\bibinfo{year}{2000}).

\bibitem{pieper01}
\bibinfo{author}{\bibnamefont{{S. C. Pieper, V. R. Pandharipande, R. B.
  Wiringa and J. Carlson }}}, \bibinfo{journal}{Phys. Rev. C}
  \textbf{\bibinfo{volume}{64}}, \bibinfo{pages}{014001}
  (\bibinfo{year}{2001});
 \bibinfo{author}{\bibnamefont{{S. C. Pieper and R. B. Wiringa}}}, 
  \bibinfo{journal}{Ann. Rev. Nucl. Part. Sci.}
  \textbf{\bibinfo{volume}{51}}, \bibinfo{pages}{53}
  (\bibinfo{year}{2001}).

\bibitem{GFMC_9_10}
\bibinfo{author}{\bibnamefont{{S. C. Pieper, K. Varga and R. B.
  Wiringa}}}, \bibinfo{journal}{Phys. Rev. C}
  \textbf{\bibinfo{volume}{66}}, \bibinfo{pages}{044310}
  (\bibinfo{year}{2002});
\bibinfo{author}{\bibnamefont{{R. B. Wiringa and S. C. Pieper}}}, 
\bibinfo{journal}{Phys. Rev. Lett.} 
\textbf{\bibinfo{volume}{89}}, \bibinfo{pages}{182501}
  (\bibinfo{year}{2002}).

\bibitem{TMprime99}
\bibinfo{author}{\bibnamefont{{S. A. Coon}}} \bibnamefont{and}
  \bibinfo{author}{\bibnamefont{{H. K. Han}}}, \bibinfo{journal}{Few-Body
  Systems} \textbf{\bibinfo{volume}{30}}, \bibinfo{pages}{131}
  (\bibinfo{year}{2001}).

\bibitem{LS1}
\bibinfo{author}{\bibfnamefont{K.}~\bibnamefont{Suzuki}} \bibnamefont{and}
  \bibinfo{author}{\bibfnamefont{S.~Y.} \bibnamefont{Lee}},
  \bibinfo{journal}{Prog. Theor. Phys.}
  \textbf{\bibinfo{volume}{64}}, \bibinfo{pages}{2091} (\bibinfo{year}{1980}).

\bibitem{LS2}
\bibinfo{author}{\bibfnamefont{K.}~\bibnamefont{Suzuki}},
  \bibinfo{journal}{Prog. Theor. Phys.} \textbf{\bibinfo{volume}{68}},
  \bibinfo{pages}{246} (\bibinfo{year}{1982}).

\bibitem{LS3}
\bibinfo{author}{\bibfnamefont{K.}~\bibnamefont{Suzuki}} \bibnamefont{and}
  \bibinfo{author}{\bibfnamefont{R.}~\bibnamefont{Okamoto}},
  \bibinfo{journal}{Prog. Theor. Phys.} \textbf{\bibinfo{volume}{70}},
  \bibinfo{pages}{439} (\bibinfo{year}{1983}).

\bibitem{UMOA}
\bibinfo{author}{\bibfnamefont{K.}~\bibnamefont{Suzuki}} \bibnamefont{and}
  \bibinfo{author}{\bibfnamefont{R.}~\bibnamefont{Okamoto}},
  \bibinfo{journal}{Prog. Theor. Phys.} \textbf{\bibinfo{volume}{92}},
  \bibinfo{pages}{1045} (\bibinfo{year}{1994}).

\bibitem{Trlifaj}
\bibinfo{author}{\bibfnamefont{L.}~\bibnamefont{Trlifaj}},
  \bibinfo{journal}{Phys. Rev. C}
  \textbf{\bibinfo{volume}{5}}(\bibinfo{number}{5}), \bibinfo{pages}{1534}
  (\bibinfo{year}{1972}).

\bibitem{MFD}
\bibinfo{author}{\bibnamefont{{J. P. Vary}}},
\bibinfo{journal}{``The Many-Fermion-Dynamics
            Shell-Model Code'', Iowa State University}
 (\bibinfo{year}{1992}) ({\bibnamefont{{unpublished}}}).

\bibitem{OrmJohn} W. E. Ormand and C. Johnson, private communication.  

\bibitem{TM}
\bibinfo{author}{\bibnamefont{{S. A. Coon, M. D. Scadron, P. C. McNamee, B. R.
  Barrett, D. W. E. Blatt and B. H. J. McKellar}}}, \bibinfo{journal}{Nucl.
  Phys. A} \textbf{\bibinfo{volume}{317}}, \bibinfo{pages}{242}
  (\bibinfo{year}{1979}).

\bibitem{Bira}
\bibinfo{author}{\bibnamefont{{J. L. Friar, D. H\"uber,
and U. van Kolck}}},
 \bibinfo{journal}{Phys. Rev. C} \textbf{\bibinfo{volume}{59}},
  \bibinfo{pages}{53} (\bibinfo{year}{1999}).

\bibitem{NCSM_6}
\bibinfo{author}{\bibnamefont{{P. Navr\'atil, J. P. Vary, W. E. Ormand 
and B. R. Barrett}}},
 \bibinfo{journal}{Phys. Rev. Lett.} \textbf{\bibinfo{volume}{87}},
  \bibinfo{pages}{172502} (\bibinfo{year}{2001}).

\bibitem{AS88}
\bibinfo{author}{\bibfnamefont{F.}~\bibnamefont{Ajzenberg-Selove}}, 
  \bibinfo{journal}{Nucl. Phys. A} \textbf{\bibinfo{volume}{490}},
  \bibinfo{pages}{1} (\bibinfo{year}{1988}).

\bibitem{Till02}
\bibinfo{author}{\bibnamefont{{D. R. Tilley, C. M. Cheves, J. L. Godwin, G. M. Hale,
 H. M. Hofmann, J. H. Kelley, C. G. Sheu and H. R. Weller}}},
  \bibinfo{journal}{Nucl. Phys. A} \textbf{\bibinfo{volume}{708}},
  \bibinfo{pages}{3} (\bibinfo{year}{2002}).

\bibitem {T88}
\bibinfo{author}{\bibfnamefont{I. Tanihata, T. Kobayashi,
                O. Yamakawa, S. Shimoura, K. Ekuni, K. Sugimoto,
                N. Takahashi, T. Shimoda and H. Sato}},
  \bibinfo{journal}{Phys. Lett. B} \textbf{\bibinfo{volume}{206}},
  \bibinfo{pages}{592} (\bibinfo{year}{1988});
\bibinfo{author}{\bibfnamefont{A. Ozawa, I. Tanihata, T. Kobayashi,
                Y. Sugahara, O. Yamakawa, K. Omata, K. Sugimoto,
                D. Olson, W. Christie and H. Wieman}},
  \bibinfo{journal}{Nucl. Phys. A} \textbf{\bibinfo{volume}{608}},
  \bibinfo{pages}{63} (\bibinfo{year}{1996});
\bibinfo{author}{\bibfnamefont{H. De Vries, C. W. De Jager and C. De Vries}},
   \bibinfo{journal}{At. Data Nucl. Data Tab.} \textbf{\bibinfo{volume}{36}},
  \bibinfo{pages}{495} (\bibinfo{year}{1987}).

\bibitem {Schia}
\bibinfo{author}{\bibfnamefont{R. Schiavilla and R. B. Wiringa}},
  \bibinfo{journal}{Phys. Rev. C} \textbf{\bibinfo{volume}{65}},
  \bibinfo{pages}{054302} (\bibinfo{year}{2002}).

\bibitem {Chou93}
\bibinfo{author}{\bibfnamefont{W.-T. Chou, E. K. Warburton and B. A. Brown}},
  \bibinfo{journal}{Phys. Rev. C} \textbf{\bibinfo{volume}{47}},
  \bibinfo{pages}{163} (\bibinfo{year}{1993}).

\bibitem{NB98}
\bibinfo{author}{\bibnamefont{{P. Navr\'atil and B. R. Barrett}}}, 
\bibinfo{journal}{Phys. Rev. C} \textbf{\bibinfo{volume}{57}},
  \bibinfo{pages}{3119} (\bibinfo{year}{1998}).

\bibitem{Be8_NCSM}
\bibinfo{author}{\bibnamefont{{E. Caurier, P. Navr\'atil, W. E. Ormand and J. P. Vary}}},
  \bibinfo{journal}{Phys. Rev. C} \textbf{\bibinfo{volume}{64}},
  \bibinfo{pages} {051301} (\bibinfo{year}{2001}).

\bibitem{A10_NCSM}
\bibinfo{author}{\bibnamefont{{E. Caurier, P. Navr\'atil, W. E. Ormand and J. P. Vary}}},
  \bibinfo{journal}{Phys. Rev. C} \textbf{\bibinfo{volume}{66}},
  \bibinfo{pages} {024314} (\bibinfo{year}{2002}).

\bibitem{Daito98}
\bibinfo{author}{\bibnamefont{{I. Daito {\it et al.}}}},
  \bibinfo{journal}{Phys. Lett. B} \textbf{\bibinfo{volume}{418}},
  \bibinfo{pages} {27} (\bibinfo{year}{1998}).

\bibitem{AS90}
\bibinfo{author}{\bibfnamefont{F.}~\bibnamefont{Ajzenberg-Selove}}, 
  \bibinfo{journal}{Nucl. Phys. A} \textbf{\bibinfo{volume}{506}},
  \bibinfo{pages}{1} (\bibinfo{year}{1990}).

\bibitem{Tadd90}
\bibinfo{author}{\bibnamefont{{T. N. Taddeucci {\it et al.}}}},
  \bibinfo{journal}{Phys. Rev. C} \textbf{\bibinfo{volume}{42}},
  \bibinfo{pages} {935} (\bibinfo{year}{1990}).

\bibitem{Arya85}
\bibinfo{author}{\bibfnamefont{R. Aryaeinejad {\it et al.}}}, 
  \bibinfo{journal}{Nucl. Phys. A} \textbf{\bibinfo{volume}{436}},
  \bibinfo{pages}{1} (\bibinfo{year}{1985}).

\bibitem{Wolt90}
\bibinfo{author}{\bibfnamefont{A. A. Wolters, A. G. M. van Hees 
  and P. W. M. Glaudemans}}, 
  \bibinfo{journal}{Phys. Rev. C} \textbf{\bibinfo{volume}{42}},
  \bibinfo{pages}{2062} (\bibinfo{year}{1990}).

\bibitem{Millener_pr}
\bibinfo{author}{\bibfnamefont{D. J. Millener}}, 
  \bibinfo{journal}{private communication}.

\bibitem{Millener}
\bibinfo{author}{\bibfnamefont{D. J. Millener}}, 
  \bibinfo{journal}{Nucl. Phys. A} \textbf{\bibinfo{volume}{693}},
  \bibinfo{pages}{394} (\bibinfo{year}{2001}).

\bibitem{c13_scat}
\bibinfo{author}{\bibnamefont{{G. Thiamov\'a, V. Burjan, J. Cejpek, V. Kroha
and P. Navr\'atil}}},
  \bibinfo{journal}{Nucl. Phys. A} \textbf{\bibinfo{volume}{697}},
  \bibinfo{pages}{25} (\bibinfo{year}{2002}).

\bibitem{c13_millener}
\bibinfo{author}{\bibfnamefont{D. J. Millener, D. I Sober, H. Crannell, 
  J. T. O'Brien, L. W. Fagg, S. Kowalski, C. F. Williamson and L. Lapik\'as}},
  \bibinfo{journal}{Phys. Rev. C} \textbf{\bibinfo{volume}{39}},
  \bibinfo{pages} {14} (\bibinfo{year}{1989}).

\bibitem{AS91}
\bibinfo{author}{\bibfnamefont{F.}~\bibnamefont{Ajzenberg-Selove}}, 
  \bibinfo{journal}{Nucl. Phys. A} \textbf{\bibinfo{volume}{523}},
  \bibinfo{pages}{1} (\bibinfo{year}{1991}).

\bibitem{CCM}
\bibinfo{author}{\bibfnamefont{B. Mihaila and J. H. Heisenberg}},
  \bibinfo{journal}{Phys. Rev. C} \textbf{\bibinfo{volume}{61}},
  \bibinfo{pages} {054309} (\bibinfo{year}{2000}).

\bibitem{Friar}
\bibinfo{author}{\bibnamefont{{S. A. Coon and J. L. Friar}}},
 \bibinfo{journal}{Phys. Rev. C} \textbf{\bibinfo{volume}{34}},
  \bibinfo{pages}{34} (\bibinfo{year}{1986}).

\bibitem{Bira1}
\bibinfo{author}{\bibnamefont{{U. Van Kolck}}},
 \bibinfo{journal}{Phys. Rev. C} \textbf{\bibinfo{volume}{49}},
  \bibinfo{pages}{2932} (\bibinfo{year}{1994}).

\bibitem{EFT_V3b}
\bibinfo{author}{\bibfnamefont{E. Epelbaum, A. Nogga, W. Gl\"ockle, H. Kamada, 
                               Ulf-G. Meissner and H. Witala}},
  \bibinfo{journal}{Phys. Rev. C} \textbf{\bibinfo{volume}{66}},
  \bibinfo{pages} {064001} (\bibinfo{year}{2002}).

\bibitem{Andreas}
\bibinfo{author}{\bibfnamefont{A. Nogga}},
  \bibinfo{journal}{private communication}.

\end{thebibliography}
\end{document}